\documentclass[useAMS,usenatbib]{mn2e}
\usepackage{psfig}

\def\apj{{\rm ApJ}}
\def\mnras{{\rm MNRAS}}

\def\etal{{\rm et~al.\ }}
\def\hmpc{\;h^{-1}{\rm Mpc}}

\def\hkpc{h^{-1}{\rm kpc}}
\def\kms{{\rm \;km\;s^{-1}}}

\def\msun{{\rm h^{-1} M_{\odot}}}

\def\simlt{\lower.5ex\hbox{$\; \buildrel < \over \sim \;$}}
\def\simgt{\lower.5ex\hbox{$\; \buildrel > \over \sim \;$}}

\title[Galaxy morphology in ${\rm \Lambda}$CDM]{
Galaxy morphology, kinematics and clustering in a
hydrodynamic simulation
of a  ${\rm \Lambda}$CDM universe
}

\author[R.A.C. Croft \etal]{
Rupert A.C. Croft$^{1,2}$\thanks{E-mail: rcroft@cmu.edu},
Tiziana Di Matteo$^{1,2}$, Volker Springel$^{3}$ and Lars Hernquist$^{4}$\\
$^{1}$ Dept.   of  Physics,   Carnegie   Mellon  University,
Pittsburgh, PA 15213, USA\\
$^{2}$ Bruce and Astrid McWilliams Center for Cosmology,
Carnegie   Mellon  University,
Pittsburgh, PA 15213, USA\\
$^{3}$ Max Planck Institute for Astrophysics,
85741 Garching, Germany\\
$^{4}$ Harvard-Smithsonian Center for Astrophysics,
 60 Garden Street, Cambridge, MA 02138, USA\\
}

\begin{document}

\pagerange{\pageref{firstpage}--\pageref{lastpage}} \pubyear{2007}

\maketitle

\label{firstpage}

\begin{abstract}
We explore galaxy properties and their link with
environment and clustering using a
population of $\sim 1000$ galaxies formed in a high resolution hydrodynamic
simulation of the ${\rm \Lambda}$CDM cosmology.  At  
the redshift we concentrate on, $z=1$, 
 the spatial resolution is 1.4 proper $\hkpc$ and the mass resolution is
such that Milky-way sized disk galaxies contain $\sim 10^{5}$
 particles within their virial radii. The simulations include
 supermassive black hole accretion and feedback as well as a multiphase
model for star formation.  
Overall, we find that a number of 
familiar qualitative relationships hold approximately
between galaxy properties, for example we observe galaxies as
lying between two broad extremes of type, where ``late'' types 
tend to be smaller in size, have lower circular velocities, younger
stars, higher star formation rates, larger disk to bulge ratios and lower
Sersic indices than ``early types''.
We find that as in previous studies 
of small numbers of resimulated galaxies the stellar component of 
disk galaxies is not as rotationally supported as in observations. Bulges
contain too much of the stellar mass, although exponential disks do have
scale lengths compatible with observations, and
the stellar mass Tully-Fisher relation at $z=1$ is reproduced.
 The addition of black hole
physics to the simulations does not appear to have made a large impact
on the angular momentum results compared to these other studies, nor do we find
that running an identical simulation with significantly lower mass resolution 
affects this aspect. Despite this, we can profitably use the rank order
of either disk to total ratio, Sersic index, or galaxy age to separate galaxies
into morphological classes and examine the density-morphology relation and
morphology dependence of clustering. We find that while at redshift $z=0$,
the well known preponderance of early types in dense environments is seen,
at $z=1$ the density-morphology relation becomes flatter and late type
galaxies are even seen to have a higher clustering amplitude than
early types.

\end{abstract}

\begin{keywords}
Cosmology: observations -- Galaxies, large-scale structure of Universe
\end{keywords}

\section{Introduction}

Forming realistic galaxies is one of the grand challenges for numerical 
cosmology. A promising approach assumes that this can be done, 
given the correct cosmological model and enough
spatial and mass resolution in a simulation of the physics of gravitational
collapse of gas and dark matter,  
Pioneering work in this vein (e.g., Evrard 1988,
 Hernquist \& Katz 1989) included some of the
 physical implementations (Smoothed Particle Hydrodynamics, 
Gingold \& Monaghan, 1977, Lucy 1977)
still most used today. Gradual improvements and additions, including 
 star formation (e.g. Katz 1992), feedback (e.g., Springel
and Hernquist 2003a), black hole accretion (e.g.,
Di Matteo \etal 2005, Springel \etal 2005, Okamoto \etal 2007) have
 led to better modelling of galaxy properties (e.g., Abadi \etal 2003a,b,
Governato \etal 2007). The most striking result to come from this research
effort is the generic problem of too little angular momentum 
and too centrally concentrated galaxies (see
e.g., the recent review by Mayer \etal 2008). Many aspects
of our understanding of galaxy formation have been progressing, however,
spurred on by high resolution ``zoomed'' resimulations of individual
galaxies (e.g. Abadi \etal 2003, Sommer-Larsen \etal 2003,
 Governato \etal 2004,  Robertson \etal. 2004, Zavala \etal 2007)
 In the meantime, larger-scale simulations (e.g., Springel
\etal 2005, Weinberg \etal 2004) have led to better predictions of the
large scale clustering properties of galaxies. In the present paper, we aim
to address questions which link both approaches,
by looking at internal properties of galaxies which form
in a cosmological volume with uniformly high resolution. In this way
we can look at not only the properties of a wide range of galaxies, 
but the relation of these properties with their large scale 
environments.

Using the same simulation to study individual galaxies and the 
evolution of structure in a cosmological volume is necessary to 
ensure that we deal with a fair sample of galaxies. In zoomed resimulations
(e.g., Katz \etal 1994, Robertson \etal 2005)
 galaxies that are chosen to be followed at high 
resolution (usually systems relatively free of major mergers at late types) 
may be special by virtue of their selection, and it is not easy to 
quantify biases that this may induce in statistical studies of their 
properties.  Running a cosmological volume 
with enough resolution to carry this out to redshift $z=0$ is
extremely expensive computationally. Motivated both by this and the
rapid growth of observational data available for redshift $z=1$, (e.g.,
the DEEP2 study of galaxy properties and environment at
$z \sim1$, Cooper \etal 2007, the VVDS survey, Meneux \etal 2008, and 
COMBO-17 survey, Bell \etal 2004) we focus most of our attention on
a large, high resolution simulation run to this redshift. Results from this 
computation, known as the {\it BHCosmo} run were first presented by 
Di Matteo \etal (2007). It features the first direct inclusion of supermassive
black hole accretion and feedback in a cosmological hydrodynamic model.

The questions which can be asked include whether we are able to make
disk galaxies which look like observed spirals, and whether they lie on the 
Tully-Fisher relation. Having a large statistical sample of simulated
galaxies will enable studies of how their properties are
correlated and interrelated, for example how disk scales lengths are related
to circular velocities and so on. In the present study we do not carry out
population synthesis modelling, instead concentrating on morphology
of stellar mass density, stellar and gas kinematics and other related
measures of galactic structure.

Observationally, large datasets of galaxies at redshifts $z\simgt 1$ are
becoming available. As mentioned above, with the DEEP2 redshift 
survey (see e.g., Coil \etal 2008 and references therein)
 data it has become possible to look at the dependence of
galaxy clustering on color, 
luminosity  and other properties.
HST images of galaxies with enough resolution for
morphological classification have extended studies
of the morphology-density relation to $z\sim0.5-1.0$
 (see e.g., the early work of
the MORPHS collaboration, Dressler \etal 1997
through Smith \etal 2005).
At higher redshifts, Peter \etal (2007)
have also measured galaxy morphologies in the environment of a rich cluster
at redshift $z=2.3$ using HST/ACS images.
With the increase in observational data available at these redshifts, it is
important to investigate the robustness of theoretical
 modelling with a variety of techniques. For example, 
Zheng \etal (2007) have fit Halo Occupation Distribution
models to the luminosity dependence of clustering in DEEP2 and Croton \etal 
(2006) have made semi-analytic model
predictions for the clustering strengths of red and blue 
galaxies at redshifts relevant for DEEP2. In the present work, we explore
the predictions of hydrodynamic simulations.

Our plan for this paper is as follows. In Section 2 we describe the 
cosmological simulations,  how we select galaxies from the
simulation outputs and show some images of examples 
of late and early type galaxies. In Section 3 we investigate galaxy properties
in detail, including kinematically defined bulge to disk ratios, 
and projected stellar density profiles. We show how measurements of these
quantities are affected by simulation resolution and then examine the
relationships between all the properties for individual
galaxies. We also plot the angular momentum content of late type
galaxies and compare it to observations, as well as comparing to
the observed stellar mass Tully-Fisher relation. In Section 4 we focus
on the relationship between galaxy properties and environment, including 
comparison to the observed density morphology relation. In Section 5 we
look at large scale structure in the simulated galaxy distribution and 
compute the autocorrelation function. We summarize and discuss our findings
in Section 6.

\section{Simulation}

We have carried out a large cosmological simulation, using the 
SPH code Gadget-2 (Springel 2005), with the inclusion of 
modelling of black hole accretion and feedback, previously
used in simulations of isolated galaxies and galaxy
mergers (Di Matteo \etal 2005,
Springel \etal 2005). This direct cosmological simulation
of black hole evolution in a cosmological context,
known as the {\it BHCosmo} simulation is presented and
described in more detail in Di Matteo \etal (2007).
The simulation algorithms and description of the 
black hole population that forms and its relationship to 
galaxies in the simulation are also set out in that paper.
A slightly different approach to modelling black holes in  cosmological
Gadget-2 simulations, including feedback from
a ``radio mode'' is also presented by Sijacki \etal (2007).
Okamoto \etal (2007) have also incorporated their own modelling of
black hole physics in zoomed simulations (also run with Gadget-2) of
individual galaxies. We discuss some similarities with our results in 
Section 6.2

Besides black hole feedback, the simulation includes the
multiphase star formation model of Springel \& Hernquist (2003a), in
which stellar wind particles can propagate from the sites
of star formation and carry energy over large distances. This
model was previously used in the zoomed simulation of an
individual disk galaxy in a cosmological context by Robertson \etal (2005).

The cosmological parameters used as inputs to the {\it BHCosmo} simulation
are consistent with the WMAP first year results (Spergel \etal 2003),
i.e. $h=0.7=$H$_{0}/100 \kms$ Mpc$^{-1}$, matter content $\Omega_{m}=0.3$,
cosmological constant  $\Omega_{\Lambda}=0.7$, baryon content
$\Omega_{b}=0.04$ and amplitude of mass fluctuations, $\sigma_{8}=0.9$.
The simulation volume is a periodic cube of side length 33.75 $\hmpc$,
and the {\it BHCosmo} contains $486^{3}$ dark matter and initially
 $486^{3}$ gas particles, giving a mass resolution of $2.75 \times
10^{7} \msun$ per
dark matter particle and $4.24 \times 10^6 \msun$
 per gas particle. We evolve the simulation
from redshift $z=99$ down to $z=1$, using constant gravitational
softening in comoving units. At $z=1$, the force resolution
is $1.4$ proper $\hkpc$.

The {\it BHCosmo} simulation has a mass and force resolution that fits
into the scheme devised by Springel \& Hernquist (2003b), who
ran a set of simulations with box size varying by factors of $\sim 3$ 
and particle numbers by factors of $1.5^{3}$. In their naming
scheme, the {\it BHCosmo} would be equivalent to a D6 simulation.
We have also run another simulation with the same initial conditions,
box size  and physical modelling but coarser mass and force resolution 
(equivalent to D4, i.e. $2\times216^{3}$ particles). We use this
D4 run, which has a mass per particle 11.4 times greater than the
{\it BHcosmo}, and force resolution 2.3 times
larger, to gauge the effects of
resolution on our results, as was done by Di Matteo \etal (2007).
Because the D4 was run with the same random initial Fourier phases as
the {\it BHcosmo} it is possible to identify the same objects which form
in both runs and therefore make direct comparisons (we shall do this
in \S 3.3).

\subsection{Galaxy selection}

We select galaxies from the particle distributions using a variant of
the 
subgroup finder {\small SUBFIND} (see Springel \etal 2001).
This selects gravitationally
bound clumps of particles, using the gas, dark matter and stellar
content of groups to do this. Each subgroup is therefore deemed
to be an individual galaxy. Each galaxy is also 
associated with a friends-of-friends group, which is found
using the usual method, linking together particles connected
by less than 0.2 times the mean interparticle separation. Particles within
the friends-of-friends group are approximately those that lie within
the virial overdensity threshold of a dark matter halo. 

\begin{figure*}
\centerline{
\psfig{file=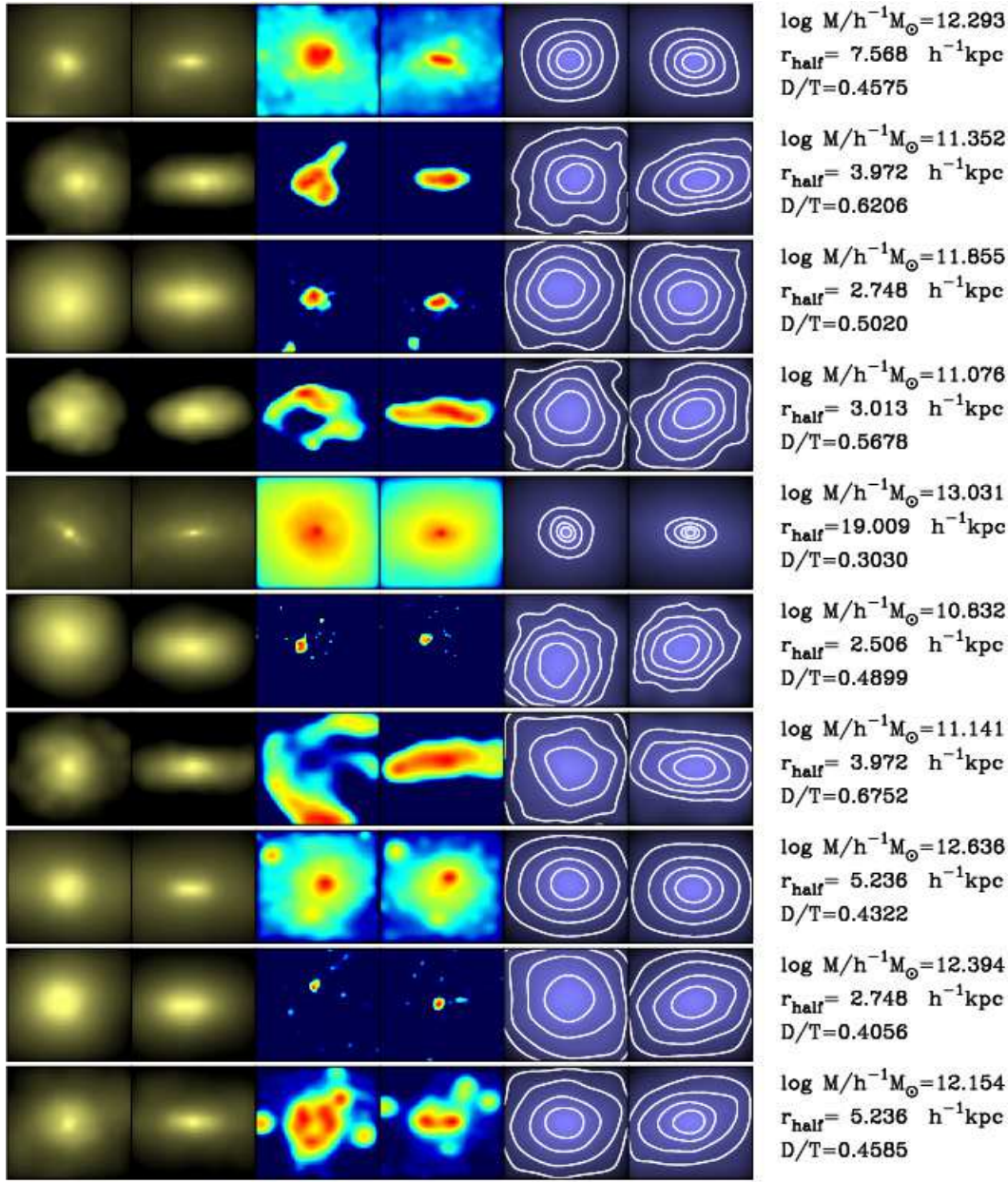,angle=-90.,width=14.0truecm}
}
\caption{
Images of $z=1$ galaxies from the {\it BHCosmo}
simulation selected kinematically to have a high D/T ratio (i.e.,
``late types'', see text \S3.1). We show 10 galaxies, in no
particular order, from the top to the bottom of the page. The
three columns with different appearances are from left to right the
stellar surface density, the gas surface density and the dark matter 
surface density in a cube centered on the minimum of the
gravitational potential for the galaxy
(see \S2.1). In each column, there are two panels corresponding
to two projections, the leftmost is such that the galaxy is face on,
using the angular momentum vector of the star particles in the galaxy
to define the direction. The rightmost panel is an edge on projection.
All surface densities are plotted with a logarithmic scale,
and the contours in the dark matter plots are shown at intervals of 0.2 dex
The length scales are such that each panel shows a region of space
from $-2 r_{\rm half}$ to $+ 2 r_{\rm half}$, where  $r_{\rm
  half}$ is the radius which encloses half the stellar mass.
The values of $r_{\rm half}$ for each galaxy (in proper $\hkpc$) are
shown to the right, along with the D/T ratio and total mass of the
galaxy/subgroup (including stars, gas and dark matter.)
\label{plotspirals}
}
\end{figure*}

From the list of galaxies, we pick those which have more than 5000 particles
in the {\it BHCosmo} run,
in order to limit the noise on morphological and kinematic properties
measured from the galaxy particle distributions. At $z=1$, we have
1180 galaxies in the {\it BHCosmo} simulation. The particle
number cut is essentially equivalent to a cut in circular 
velocity of $100 \kms$. We find that at $z=1$, galaxies  
of similar mass to the Milky Way (those with circular velocities within 
$10\%$ of $220 \kms$) contain $\sim 60000$ particles
 on average.  For the D4 low resolution run we use an appropriately
lower threshold on the particle number (439 particles).

For each galaxy (i.e. subgroup) we define a central point, which is
the particle (star, gas or dark matter) at the minimum of the
gravitational potential.
This particle serves as a center when computing the stellar density and
angular momentum profiles and for all other purposes in the paper 
when the center of the galaxy is required.

\subsection{Example galaxies}

\begin{figure*}
\centerline{
\psfig{file=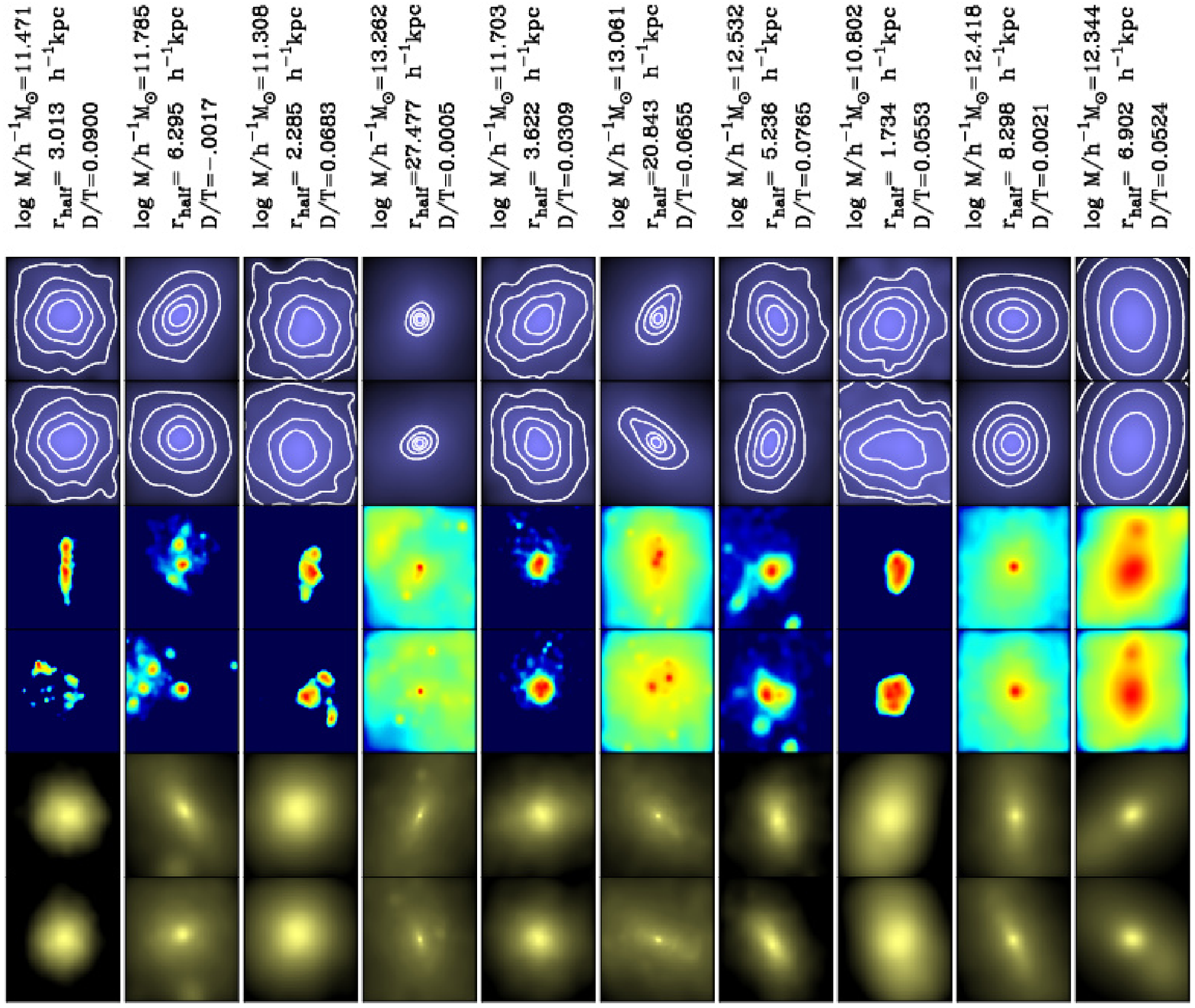,angle=-90.,width=14.0truecm}
}
\caption{
Images of ten $z=1$ galaxies from the {\it BHCosmo}
simulation selected kinematically to have a low D/T ratio (i.e.,
``early types'', see text \S3.1). We show the stellar, gas and dark matter
surface density from left to right, in the same fashion as the
late type galaxies in Figure \ref{plotspirals}.
\label{plotellip}
}
\end{figure*}

Before characterizing the statistical properties of our sample
of galaxies, we examine their morphology visually. To do this,
we take galaxies from the redshift $z=1$ output of the
{\it BHCosmo} run. In Figures
\ref{plotspirals}
and \ref{plotellip},  we show 10 representative galaxies taken from
two classes, those with high D/T (Disk to Total)
ratios and low D/T ratios respectively. The D/T ratio is derived from 
a kinematic decomposition of the stellar mass content of each galaxy
(following e.g., Abadi \etal 2003a), and
will be more fully explored in \S3.1 For now it is sufficient to state that
these decompositions form the basis for one possible separation into ``late''
type disky galaxies and ``early'' type ellipticals. 

Figure \ref{plotspirals} shows galaxies that lie principally in the
top 10\% or so of D/T values, although because most galaxies are 
bulge dominated, even these include some with D/T values less than 0.5.
We note that zoomed resimulations of disk galaxies in other work
typically have similar D/T ratios (see e.g.,
Abadi \etal 2003b in which the galaxy had D/B of 0.83, or D/T of 0.45).
The plot shows two views each of the stellar density, gas density and
dark matter density, on a logarithmic scale. If the angular momentum
vector of the stellar component defines the $z$-axis,
the two panels for each component show the face on ($x-y$)
view and a side on ($x-z$) view. We can see that the stellar components
are flattened in the direction expected if they are rotationally supported.
They are not particularly thin, and an obvious bulge is not
easy to spot in the surface brightness plots. Comparing with the
zoomed simulations of disk galaxies performed by Robertson
\etal 2004 (their figure 3, top left) or Governato \etal 2007 (their figure 6,
left panels),
we can see similar morphologies. 

The gas components in  Figure \ref{plotspirals}
are more varied in appearance than the stars, with most galaxies 
having relatively thin disks. Some however
(such as the second and third from bottom) have rather 
clumpy distributions and are less obviously flat.
Approximately half the dark matter halos
can be seen to be obviously
 slightly flattened in the same direction as the stellar disks.
Alongside the plots, we have listed the D/T ratios and the total mass of the
subhalo. Most of the disk galaxies we plot are fairly large, with total
mass around $10^{12} \msun$. The fifth from the top, the largest and
most massive has the lowest D/T ratio. We also show next to each panel
an indicator of the physical size of the galaxy, the radius
of a sphere which encloses half the stellar mass ($r_{\rm half}$),
in proper $\hkpc$.

Moving to the early types in Figure \ref{plotellip}, one immediate 
difference that can be seen from the stellar density plots is that they
do appear to be more centrally concentrated than the disks. On average,
the ratio of major to minor axes appears to be lower, and
 the orientation of the galaxy also appears to have little relation
to the angular momentum direction (as expected, give that these
galaxies were selected on the basis of their stellar kinematics).
The gas components are almost all rather clumpy and roughly spherically
distributed, although the first galaxy plotted has a very obvious thin
gas disk.  This is rather unusual, given the almost spherical nature
of its stellar distribution, although we note that it does have the
highest D/T ratio of all the galaxies in this plot.
Interestingly, the dark matter contours
do not appear to be oriented in the same way as the stellar distribution,
as any 
visible degree of correlation between the major axes of the dark matter 
and stars is hard to pick up on. The gas distributions do appear perhaps
 to line up slightly better with the dark matter. We 
have however computed the
median angular difference between the directions of the principal axes
of the stars and gas (we return to this in \S3.7 below) and find
a value of 24 $^\circ$, a fairly strong alignment. This value is for all 
galaxies though, and presumably strongly influenced by the
good alignments of gas and stellar disks.

\section{Galaxy properties}

We will focus on three main properties of our simulated galaxy sample,
their kinematics (angular momentum and kinematic decomposition into
disk and bulge), their profiles (quantified by a Sersic index) and their
mean stellar ages and star formation rates. We do not classify
their morphology by eye as has been done observationally (e.g., by
the MORPHS group, Dressler \etal 1997 ),
but we will use the kinematic decomposition and Sersic indices
as indicators of morphology, for example when investigating the
morphology-density relation. We make use of our two simulations,
the {\it BHCosmo} run and the D4 run which have identical 
initial conditions but different mass and spatial resolution to 
investigate the effects of biases in galaxy properties due to
numerical effects. Comparisons of both simulations are presented in detail
in \S3.3 below, including some quantification of the effects 
of resolution. We first explain the measures of kinematics
and stellar profiles, also including examples drawn from both simulations.

\subsection{Kinematics and D/T ratios}

We examine the stellar orbits in the galaxies, following
a technique often used in the literature (e.g., Abadi \etal  2003b),
 we use the angular momenta of individual stars to dynamically
decompose the galaxies into disks and spheroids. 

In the disk galaxies
(Figure \ref{plotspirals}), the total angular momentum vector of the
stars defines
a direction which is visually perpendicular to the stellar disk. We will 
investigate how well this holds quantitatively later (\S 3.7), but for now
we use this  direction to define the $z-$axis. 
For a star particle,
we compute the specific angular momentum in the $z$ direction, $J$ and also
its specific binding energy $E$. Following exactly the procedure
of Abadi \etal  (2003b), we compute $E$ relative to the total mass of the
galaxy within the viral radius, $r_{200}$. For each star particle we
then compute the specific angular momentum of the corotating 
circular orbit with the same binding energy, $J_{c}$. The ratio $J/J_{c}$
gives for each particle a measure of the orbital circularity. The maximum
value, $J/J_{c}=1$ occurs for particles which are following in circular
orbits in the same direction as the majority of the stars. Particles
with close to this value are therefore likely to be in a disk component.

\begin{figure}
\centerline{
\psfig{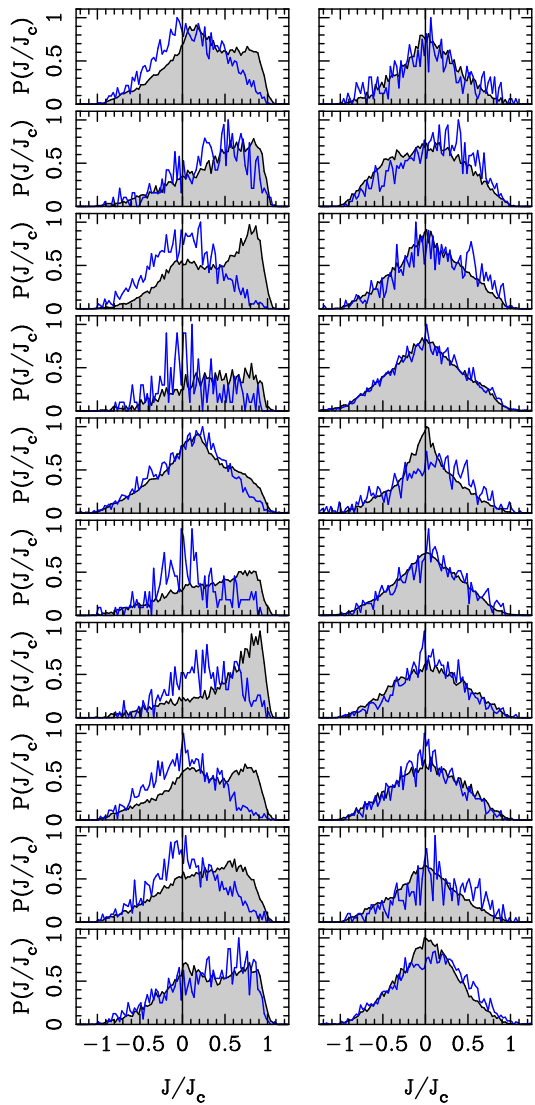}
}
\caption{
Histograms of the orbital circularity of star particles, $J/J_{C}$ 
for the galaxies (at $z=1$)
shown in Figures \ref{plotspirals} (late types, left column) and
 \ref{plotellip} (early types, right column). The quantity  $J/J_{C}$ 
is the ratio of the orbital angular momentum of an
individual  star particle to one with the same energy but on a circular
orbit (see text).
The shaded histograms are for galaxies selected from
the {\it BHCosmo} simulation and the superimposed lines are the same
galaxies taken from the lower resolution simulation, D4. How the matching
galaxies were found in the D4 run is described in \S 3.2.
The $y$-axis in each case is the relative probability density,
 normalized to 1 at the maximum value.
\label{jhist}
}
\end{figure}

In Figure \ref{jhist},
we show histograms of the $J/J_{c}$ values for all stars for our 20 example
galaxies, with the late types on the left, and early types on the right.
In order to avoid going out very far into the outskirts of the galaxy 
where particle numbers are sparse, we only plot particles which lie 
within 4 half stellar mass radii ($r_{\rm half}$) of the center of the galaxy
(the results are insensitive to this choice). To compute the D/T ratio,
we make the (somewhat arbitrary) 
assumption that the bulge has zero total angular momentum
(see e.g., van den Bosch \etal 2002). Starting from the negative
 end of the $J/J_{c}$ distribution we associate all the star particles
with negative $J/J_{c}$ with the bulge as well as an equal mass of
star particles with the
equivalent positive $J/J_{c}$ values. The star particles left over
then make up the disk. In this way, we compute 
the disk to total ratio, D/T already enumerated next to the galaxy plots
in Figures \ref{plotspirals} and \ref{plotellip}.

Because the bulge/disk decomposition relies on this $J/J_{c}$ distribution
it is not surprising that the galaxies in the left panel, chosen to be
more disky have asymmetric histograms. Eight of the ten disk galaxies
have the highest point in the $J/J_{c}$ distribution associated with
the region around $J/J_{c}=1$, showing that the galaxies indeed
have a strong stellar rotationally supported
 component, as we would expect from their morphology. Some of the 
distributions are markedly more asymetric than others, with 2 not showing 
any real evidence for a spheroid component (a bump centered on  $J/J_{c}=0$).
The shaded histogram is the result for our high resolution {\it BHCosmo}
simulation, and the (somewhat noisier) line is the result for the
same galaxies taken from the low resolution D4 model (see
Section 3.3 for details on the matching of galaxies in the simulations).
We can see that the two simulations can give rather different
results for the  $J/J_{c}$ distribution for the same galaxy. This is
likely to be due among other things
 to stochasticity coming from initial fluctuation
power below the Nyquist limit of the lower resolution model, 
and differences
in the numerical integration of particle orbits coming from 
resolution effects. This will lead to a scatter in D/T values between
the two simulations. We investigate quantitatively the effect of
resolution on this parameter below.

The low D/T galaxies in the right panel of Figure \ref{jhist}
unsurprisingly have central peaks in the  $J/J_{c}$ distribution. The
effects of resolution on the histograms themselves appears to be somewhat 
lower, and in fact we shall see that in fact the scatter in D/T
values from the two simulations is closer to being constant
in log D/T. Interestingly, if we search for evidence of a small stellar disk
to go along with the gas disk seen in the visual inspection
of the first galaxy, we see that there is no evidence for one
in the $J/J_{c}$ distribution.

Although we do not plot   $J/J_{c}$ for the gas particles, we have examined
their distribution and find it to be more rotationally supported than
the stars, in agreement with previous work (see 
e.g., Mayer \etal 2008). The previous
work on resimulated spiral galaxies has yielded D/T values
from kinematic decomposition of e.g.,
0.41 (Abadi \etal
2003b), 0.67-0.76 (Okamoto \etal 2007) 0.74 
(Governato \etal 2004). Our example disk galaxies
shown here are generally consistent with this range. From our large
sample, we can say something about how rare truly disk dominated galaxies
are in the simulation. At $z=1$ only $3.5\%$ of the galaxies
in the {\it BHCosmo} run have D/T greater than 0.5. The maximum D/T
value we find is 0.76 (a D/B ratio of 3.22). 
This is the same as the (mass weighted) D/T ratio of the disk galaxy
simulated by Okamoto \etal (2007), which included AGN feedback.

Observationally, spiral 
galaxies in the local Universe are seen to have D/T ratios from unity
(bulgeless disks) to $\sim 0.1$ (see for example Graham \etal 2001, who
find a median D/T value of 0.75 for a sample of 86 face-on spirals).
 The D/T ratios are usually (and in the case of Graham \etal)
 calculated in terms of
luminosity of each component, using fits to the photometry, rather than
in stellar mass from the kinematics, as we have done. This will tend
to make  our D/T rather lower because of the relative mass/light ratio of
disks. Nevertheless, it is clear that the disk components of our
galaxies are small, a fact related to the angular momentum
problem in simulations. In separating galaxies into early and late types,
we will therefore use a rather low value of D/T to divide the two classes
(we use D/T=0.2.)

\subsection{Projected stellar density profiles}

There are a number of ways
to decide on galaxy type using fits to the radially averaged luminosity
profiles of galaxies.
Galaxy classification based on photometry
often makes use of the fact that early type galaxies
are more centrally concentrated. 
 For example, elliptical galaxies and the bulges
of spirals tend to obey the deVaucouleurs profile well: $\Sigma\propto
\exp{(r/r_{0})^{-1/4}}$, and disk components have pure exponential profiles,
 $\Sigma\propto\exp{-r/r_{\rm eff}}$ (where
$\Sigma$ is the projected stellar density). A common method for using this
information is to fit a Sersic  profile which can account for
both,
\begin{equation}
\Sigma=
\exp{(r/r_{0})^{-1/n}}
\end{equation}
where $n$ is the Sersic index (Sersic, 1968).
At redshift $z=0$ the largest study of Sersic
index fitting was carried out by Blanton \etal (2005). At $z\sim1$, there
are quantitative
measurements of the morphology from a variety of surveys including
the CFHT legacy survey (Nuijten \etal 2005,, using the Sersic index)
 COSMOS
(e.g., Capak \etal 2007, using the Gini parameter).

\begin{figure}
\centerline{
\psfig{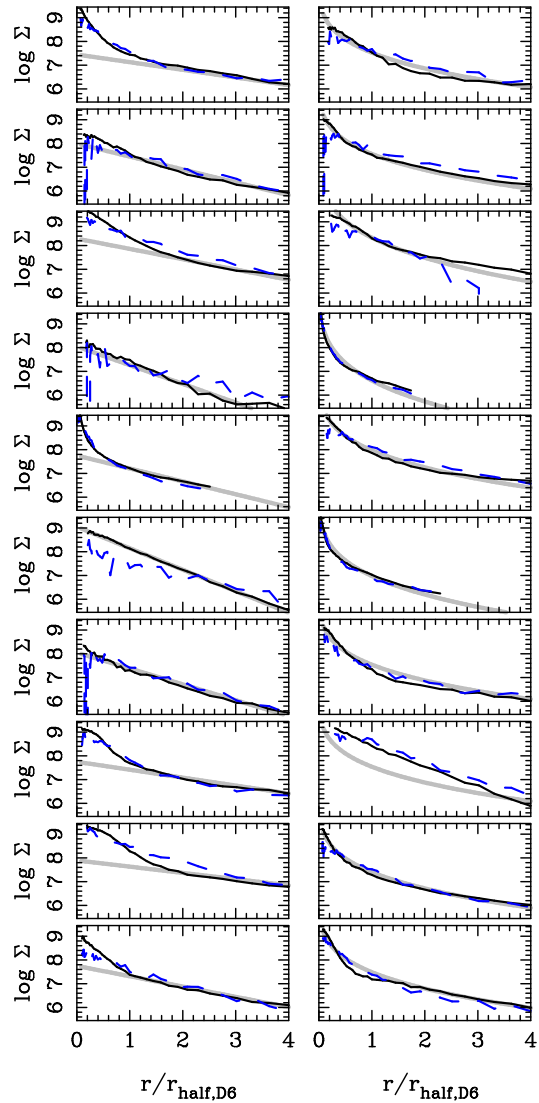}
}
\caption{
Projected stellar density profiles of the galaxies
shown in Figure \ref{plotspirals} (left column, late types)
and Figure \ref{plotellip} (right column, early types). The thin solid
lines are the profiles from the galaxies in the 
{\it BHCosmo} simulation and the
dashed lines are the profiles for the same galaxies taken from the
lower resolution D4 simulation.
The units for $\Sigma$, the projected stellar density are $\msun$ (proper
kpc)$^{-1}$.
The $x$-axis is radial distance from the center of the galaxy, as
a fraction of the half stellar mass radius (given for each galaxy
in Figures \ref{plotspirals} and \ref{plotellip}.
In the left panels we also show as thick grey lines exponential profile
fits to the profiles (beyond the $r_{\rm half}$ radius)
of the late type galaxies. In the right hand panels
we show, also as thick grey lines de Vaucouleurs profile fits to the
early-type galaxies.
\label{starprofiles}
}
\end{figure}

To examine the stellar profiles of  our simulated galaxies, we  rotate them
so that they are face on with respect to the stellar angular momentum
vector. We then use this projected distribution
to  bin the stellar particles, assigning their masses to 
radial bins. Just as with the kinematics above, we are using stellar density
rather than luminosity in a particular band. 

In Figure \ref{starprofiles} we show the projected stellar density profiles
for our 20 example galaxies, again in the same
order from top to bottom and with the late types on the left. We show results
for the D4 run as a dashed line as well as the {\it BHCosmo} run in solid.
The differences between the profiles for the two simulations appear to 
be much smaller than those in the kinematic decompositions in Figure 
\ref{jhist}. However, we shall see later that these small differences
actually lead to large scatter in the best fit Sersic index for each galaxy 
in the two simulations.

In the left panels of Figure \ref{starprofiles}  we have
also plotted an exponential disk profile, fitting it to the radial 
bins at greater distances than $r_{\rm half}$ from the center of the 
galaxy. We ignore the central bulge region within $r_{\rm half}$ in 
order to void biasing the fit, and also to avoid fitting the central profile
inside the gravitational softening length of the simulation (1.4 proper
$\hkpc$ for the {\it BHCosmo}). Looking at how well the late type galaxies
obey a pure exponential profile, we can see that 4 out of the 10 have
no significant bulge component, with the 6th from top being the best.
 A galaxy which is pure exponential
all the way to the center was seen in the resimulation of Robertson
\etal (2004). Here we see that this can happen relatively often, even with 
the comparatively low resolution of the D4 simulation.

In most of the panels in Figure \ref{starprofiles} we can see that in the
inner regions, the D4 simulation falls somewhat below the
higher resolution run. As found for example in the convergence tests
of Governato \etal (2007), the inner parts of the galaxies are
not resolved, so that we cannot say much about the structure of the bulges.
Governato \etal (2007) estimate that several million resolution elements
will be needed within the virial radius to do this.

Looking at the early types in the right panel, we can
see that they are on average more centrally concentrated, as we
noticed from looking at the images (Figure \ref{plotellip}.)
We have plotted a de Vaucouleurs profile fit to the profiles in these
images, and we can see that for  9 out of the 10 early types it gives
reasonable results.
We also 
see however that one of them (the 3rd from bottom) is also pretty close to 
exponential in shape, and the de Vaucouleurs profile is not a good fit at all.
Returning to the image of this galaxy in Figure \ref{plotellip}, it can
perhaps be seen to  be fairly smooth in the center, but there is
no real hint that its profile is peculiar, nor is there in the kinematics
Figure \ref{jhist}). This is quite suprising, and shows that even a
good exponential profile is not necessarily a sign of a disk dominated
galaxy.

In section 3.4, we shall instead fit a Sersic profile to all the galaxies,
and use that to make a separation of galaxies into two classes. For now
we have seen that for most of our 20 example galaxies there is reasonable
 agreement between the type suggested by their kinematics and their
stellar profiles.

\subsection{Resolution tests}

\begin{figure*}
\centerline{
\psfig{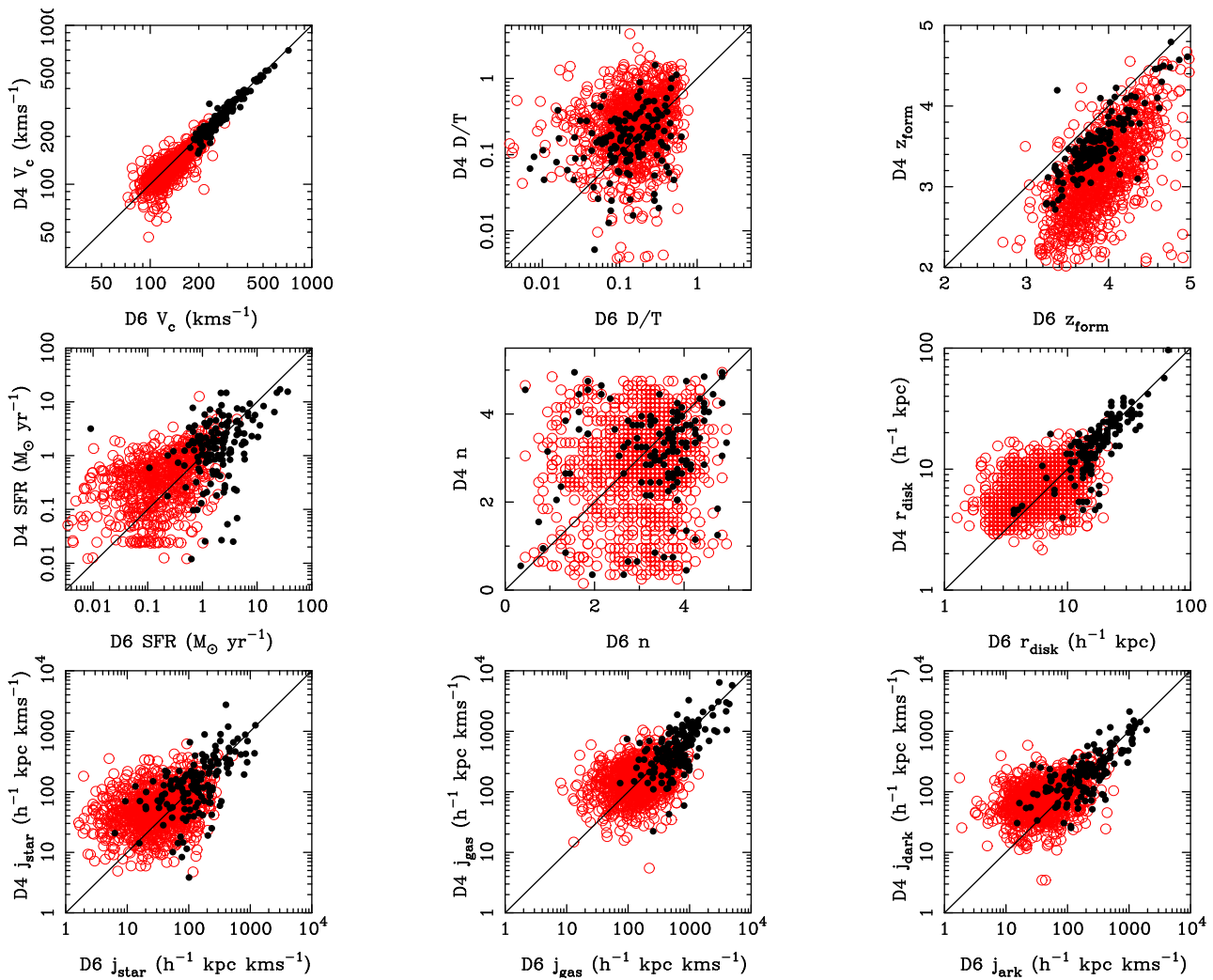}
}
\caption{
Comparison of galaxy properties in the {\it BHcosmo} (referred
to here as the D6) simulation and the same galaxies selected from
the lower resolution D4 run (see \S 3.3 for details
of how the matching galaxies were found). In each panel, the open circles
show galaxies which contain $>5000$ particles in the D6 run,
which corresponds to an equivalent of $>439$ particles in the D4 run.
The filled points are for galaxies which contain $>5000$ particles
in the D4 run (or the equivalent of $57000$ for the D6). In the top
row, we show from left to right the circular velocities of galaxies
(measured at the 2.2 times
exponential scale radius $r_{\rm disk}$ of the galaxies),
the kinematically determined D/T ratio, the mean formation redshift of
star particles in the galaxy. In the middle row we show the star formation
rate, the Sersic index, and the exponential scale radius. In the bottom
row are plotted the mean specific angular momenta of the stars,
the gas and the dark matter.
\label{matchmorph}
}
\end{figure*}

We have mentioned two of the galaxy properties we are most interested
in above, the D/T ratio and the Sersic index, $n$. We will also study the
circular velocity, $V_{\rm circ}$, the total angular momentum of 
different galaxy components (e.g., $j_{\rm gas}$), the star formation
rate, mean star formation redshift ($z_{\rm form}$) and the 
exponential disk scale length, $r_{disk}$. For all these quantities, the
numerical resolution of the simulation could influence the results. In this
section we investigate whether changing the resolution causes a systematic
bias by comparing the same galaxies in the {\it BHcosmo} and D4 runs.
Because these simulations were run with the same intitial random
seeds to set up their initial conditions and have the same box size,
in principle we should be able to find high and low resolution versions
of the same galaxies. With such a resolution test, we will be able to get
information on which of the parameters is most reliably computed by the 
simulation, in terms of both systematic bias with resolution and 
amount of scatter in results between the two resolutions.
In order to emphasise the resolution difference between the two 
simulations, we refer to the  {\it BHcosmo} run as the D6 run 
in this section (using the nomenclature of Springel \& Hernquist 2003).

In order to carry out this test we first need to match up the galaxies
in the D4 and D6 runs. We first apply a particle number cut which is
equivalent to the 5000 particle lower limit for the D6 run. After applying
this cut (439) particles we are left with 1297 D4 galaxies. For each 
D6 galaxy, we look for the closest (the smallest 3 dimensional
separation between the centers, defined by the particle at the minimum
potential) one in the D4 output. We also apply another criterion, so that 
the galaxies must be within 25 \% of each other in total mass to be 
considered a match. We do not exclude galaxies which have already been
matched. The procedure is insensitive to this, as is it is to changing the
25\% matching criterion between 10\% and 30\%. In order to gauge roughly
how many galaxies have been mismatched, we try applying the same
mass similarity
criterion to the dark matter masses of the galaxies. We find that 
none of the top 20 galaxies by mass change, but below that approximately
10\% of the D6 galaxies are matched to another in the D4. We find similar
results with the stellar mass, and so our conclusion is that approximately 10\%
of the galaxies are likely to
 be mismatched. This is a small number, but should be
borne in mind when considering the comparisons between simulations.

In Figure \ref{matchmorph}, we show the results for 9 different quantities,
plotting the value for the D4 simulation against that in the D6 simulation,
for each of the 1180 galaxies. In the plots we show two categories of 
galaxy, those which have $> 439$ particles in the D4 (equivalent
to 5000 in the D6) as open symbols and those which have 
$> 5000$ particles in the D4 (equivalent to 57000 in the D6) as filled points.
This is so that we can see whether any scatter or bias is less for larger
particle numbers per galaxy. As the D6 galaxies all have $> 5000$ particles,
in this sense the filled points will give a sense of the 
uncertainties induced by resolution on the results in the rest of the paper,
which use the D6 outputs. 

Looking at the different panels of Figure \ref{matchmorph} we can see
that all the galaxy parameters show correlations between the two simulations,
although the scatter between them is in some cases (e.g., Sersic index, $n$)
much larger than others, (e.g., $V_{\rm circ}$). An $x=y$ line has been drawn 
on all panels and the first impression is that while the scatter may be 
large, there is not much evidence of strong systematic bias caused
by decreased resolution. The most obviously different slope from 1 is that in 
the $z_{\rm form}$ panel.  Here we can see that galaxies in the
D4 simulation tend to form their stars
on average  significantly later than in the D6. For example, stars
which formed at average $z=3.5$ in the higher resolution model form at
average $z=2.5$
in the D4 run (for galaxies with between 439 and 5000 particles in the
D4). The effect of the bias is seen to be significantly reduced
when we move to galaxies with higher numbers of particles (the filled
points).  This bias with resolution is expected, given that at 
higher resolution smaller scale density fluctuations are present which 
can collapse earlier. A detailed  study of star formation 
in the same multiphase model used in the present simulation
 was carried out by Springel \& Hernquist (2003a) and to which we refer
the reader for further resolution tests and interpretation.

Related to the star formation history is the current SFR at $z=1$, which is
also shown in Figure \ref{matchmorph}. in this case we can see that the SFR in
the D4 run is systematically higher (the points tend to lie above
the $x=y$ line) than for the D6 galaxies. This is particularly true
for the low particle number galaxies.  A possible interpretation is that
perturbations which were not present in the D4 compared
to D6 due to insufficient resolution are overtaken by collapse on larger scales
and forming stars.

The two parameters which have the most scatter are the D/T ratio and the
Sersic index $n$.  The strength of the correlation between
the two simulations can be judged using the Pearson correlation
coefficient, which for D/T is 0.22 for all galaxies and 0.35 for
D4 galaxies with $> 5000$ particles. The significance levels for the
correlations are $10^{-14}$ and $2\times10^{-5}$ respectively (there 
are $\sim10$ times fewer galaxies for the high particle number 
subset, so the significance is lower). If we fit a straight line
$y=a+bx$ to the D/T results we find an offset of $a=0.21,0.11$ for
the small and large galaxies, respectively. This means that there
is a small systematic offset, but it is in the direction of the 
low resolution galaxies having higher D/T ratios, not what one would expect
if simple 
lack of resolution causes the angular momentum problem. This fit assumes
that all the errors lie in the $y-$coordinate (the D4 D/T ratio) and gives
 slopes of $b=0.54$ and $b=0.60$. The fact that the slope is not equal to one
may be partly due to a significant  portion of the scatter not
coming just from the $y-$coordinate (i.e., the D6 value is
inaccurate) and partly due to resolution
lowering the angular momentum. Because the relevant ``errors'' in the
$x$ and $y$-coordinates are not known, it is not possible to say more than
this.

For the Sersic index, we have a significantly worse correlation for 
all galaxies, $r=0.11$, with a significance of $8\times10^{-5}$.
For the $>5000$ particle galaxies it is surprisingly good, $r=0.84$ 
(significance $ < 10^{-15}$). Visually, the scatter does appear to be
very large, something that is interesting given the good correspondence
between the lines showing the stellar profiles in the D4 and D6 cases
in our plots of example galaxies (Figure \ref{starprofiles}). This shows that
differences between the slopes of the profiles are quite subtle and
that $n$ may be a problematic indicator of galaxy type.
Given the evidence of the profiles
seen in Figure \ref{starprofiles}, the
  physical gravitional softening length difference 
between the two simulations,
 (1.4 $\hkpc$ vs 3.1 $\hkpc$) does not seem to make a difference
to the Sersic indices.

\begin{figure*}
\centerline{
\psfig{file=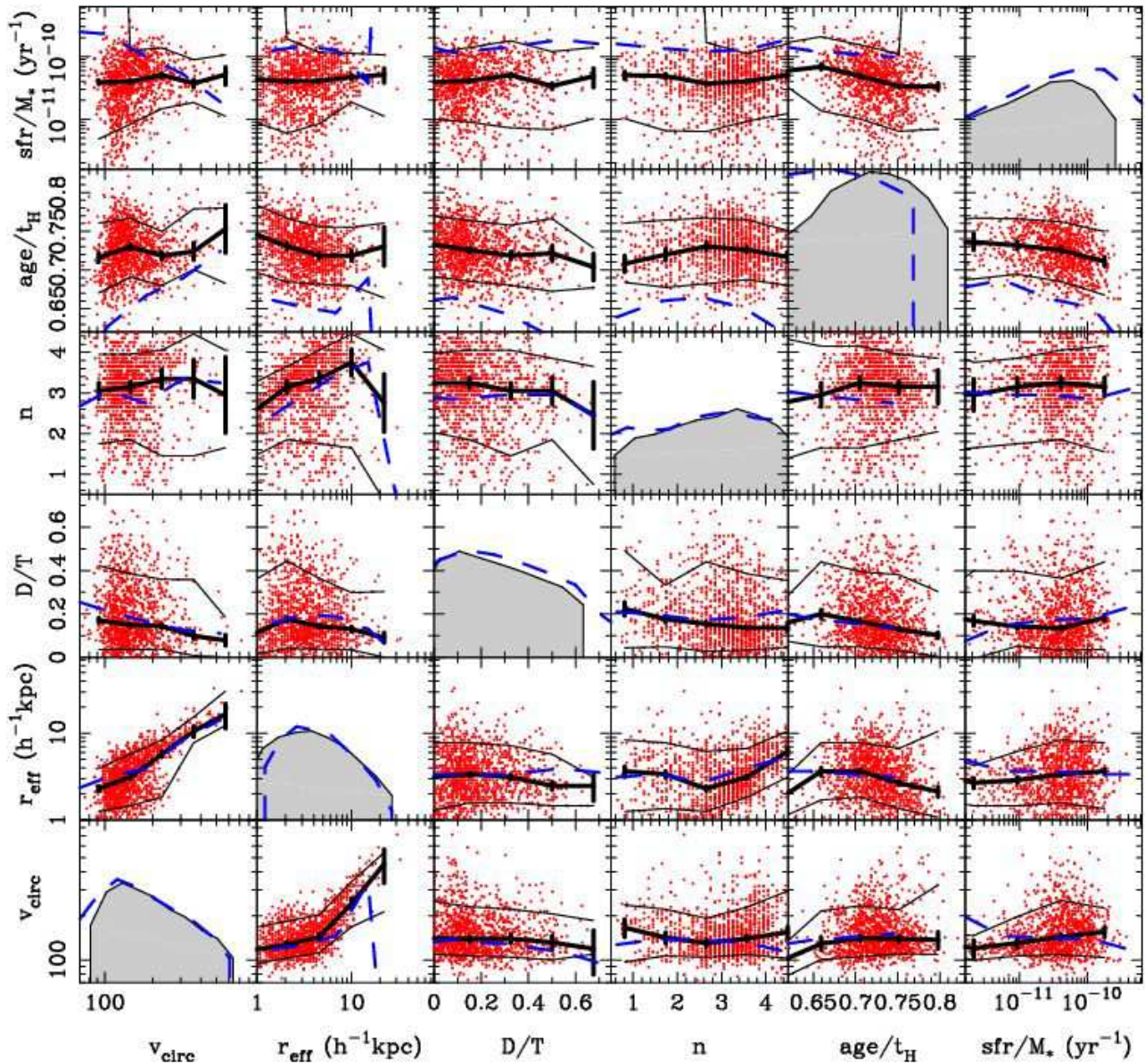,angle=-90.,width=17.0truecm}
}
\caption{
Galaxy properties in the {\it BHCosmo} simulation at redshift $z=1$.
In the panels which contain scatter plots, we show 1 point for each 
galaxy. We also show the median value in bins as a thick solid line,
with Poisson error bars. The 90th and 10th percentile distibutions in the
same bins as the median are also shown as thin solid lines. The thick dashed 
line in each case shows results for the galaxies taken from the lower
resolution D4 simulation. The diagonal panels show the probability
distribution of the parameter values, on the same $x$-axis scale as the
scatter plots. Again the dashed lines show the pdf from the D4 run and the
solid line enclosing a shaded region the results from the {\it BHCosmo}
simulation. The $y$-axis of the diagonal panels represents a logarithmic
scale of probability. The overall normalization is arbitary, but the 
height of each panel is 4 dex.
\label{36pan}
}
\end{figure*}

The bottom three panels of Figure \ref{matchmorph} show the
resolution comparison of the specific angular
momentum $j$ of each of the three components,
gas, dark matter and stars, of each galaxy.
The scatter is relatively large, worse visually
than the impression
produced by $V_{\rm circ}$ and by $r_{\rm disk}$. It does decrease with
higher particle number and there does not appear to be a
significant bias with resolution. This may mean that increasing 
resolution is not the whole story for dealing with the angular
momentum (see e.g., Agertz \etal 2007) or else that with
much higher resolution we might suddenly see a dramatic change
(see e.g., Ceverino \& Klypin 2007 for exploration along
these lines) .
 One fact which is obvious is that the
gas has higher $j$ values than either the stars or the dark matter.
We shall return to this in Section 3.5 below.

The conclusion we can draw from our study of identical galaxies at 
lower resolution is that most of the properties are robust, although the
scatter between two resultions can be large. It is largest for
the Sersic index $n$ and D/T ratios, two of the parameters we will 
use to separate our galaxies into different types. A
relatively large systematic bias between
parameters due to resolution effects is only present for mean stellar age,
which is affected in the way one expects due to missing small scale power.

\subsection{Correlations between galaxy properties}

One of the aims of this paper is to see whether galaxy morphology
in simulations is affected by environment. The other main aim is to
see how galaxy properties are interrelated. For example, we look at 
whether disk dominated galaxies are smaller than bulge dominated ones,
how Sersic index is related to D/T ratio and so on. Our main tool in this
regard is a scatter plot of 6 important galaxy properties where each is
plotted against the 5 others. This is shown, for $z=1$, the redshift
we are focussing on (for the {\it BHCosmo} run) in Figure \ref{36pan}.
For each of the quantities we show a histogram of the values on 
the diagonal (where the $y-$axis is in log units).

We have seen in Figure \ref{starprofiles} that it is possible to
roughly fit an exponential profile
 to the region outside the inner half stellar
mass radius, even when the galaxy is an early type. In this way, we have
computed an effective radius $r_{\rm eff}$ for all galaxies, so that 
we can compare early and late types on the same footing. We use this
value in our measurement of the galaxy circular velocity,
which we compute using 
\begin{equation}
V_{\rm circ}=\sqrt{GM_{<2.2r_{\rm eff } }}/r_{\rm eff}.
\label{vce}
\end{equation}
Here $M_{< 2.2r_{\rm eff}}$ is the total mass within 2.2  effective radii 
(e.g., Courteau 1997). Our $V_{\rm circ}$ values for all galaxies are therefore
computed in this way. We use $ r_{\rm eff}$ in order to 
be consistent with observational definitions of circular velocity,
but find no significant difference to our results if we instead
use $r_{\rm half}$. 
 
In the panels of Figure \ref{36pan}, we show lines which represent the
medians of the points in bins, as well as the 10th and 90th percentile lines.
For the median, as well as the pdfs, we show results for the D4 run
as a dashed line. We note that half of the panels are repeated, in a different
orientation. We include them because seeing the median of the other
quantity is still useful.

A number of interesting facts are apparent from a look at
Figure \ref{36pan}. The narrowest relation is between $r_{\rm eff}$
and $V_{\rm circ}$. Galaxies with 
 $V_{\rm circ}=200 \kms$ have a median $r_{\rm eff}$ of $4$ proper  $\hkpc$,
or 5.7 kpc. This can be compared to the Milky way, which has
$r_{\rm eff}=3.5$ kpc, for example. It seems that the size of galaxies
in the simulation is therefore roughly correct, at 
least within a factor of 2. The main problem, as we
shall see from the D/T ratios is that they have bulges which are too large.
The values of $r_{\rm eff}$ and  $V_{\rm circ}$ can be seen from 
Figure \ref{36pan} to be unaffected by resolution.

Moving on to the D/T ratios, we can see that there is a trend for disk
dominated galaxies to be smaller (lower $V_{\rm circ}$ and
$r_{\rm eff}$ values) than systems with greater bulge components. The median 
$D/T$ value for the smallest galaxies plotted ($V_{\rm circ} \sim 100 \kms$)
is approximately twice that of the largest ($V_{\rm circ} \sim 500 \kms$).
This trend can also be seen in the 90\% envelope, and
the  10\% envelope, which both trend downwards with
galaxy size. There is a particularly 
obvious lack of disk galaxies above $V_{\rm circ} \sim 300 \kms$.
Even at $z=1$, therefore the simulations exhibit the well known observational
trend for the largest galaxies to be early types.
While the trend of  D/T  with size does seem to be reasonable
compared to observations, the absolute values are low, as we have seen before.
For example, there is a lack of disk galaxies which have no detectable bulge,
but which are known to exist in the local Universe (e.g.,
Dalcanton \& Bernstein 2000).

If we look at the age of galaxies next, it becomes
 apparent that the behaviour of galaxy property trends in the presence
of large scatter is quite complex. For example in the panel of $V_{\rm circ}$
vs age, the higher  $V_{\rm circ}$ galaxies are slightly older in the
median, although the 
errors are large. On the other hand, in $r_{\rm eff}$ vs age, the smaller
galaxies (these with $r_{\rm eff} < 5 \hkpc$) have higher ages. Conversely,
the lower D/T galaxies also have higher ages, as we would expect
of late types. We find therefore that some of the usually assumed
 relationships between galaxy properties and type (see e.g.,
Maller \etal 2006, Koda \etal 2007) hold, but 
in the grid there are several counterexamples.
The effect of resolution on galaxy ages is for them to be lower in the 
D4 run, as we have seen in Section 3.3. This effect is shown by the median
line in the panels of Figure \ref{36pan}. We do see however that although 
the ages are lowered by a constant offset, the trend of the ages
with other properties is broadly similar to the higher
resoution run in all the panels.

The Sersic index, $n$, of the galaxies can be seen to increase
with  $r_{\rm eff}$ (although the last  $r_{\rm eff}$
bins drops down, the errors are large). Diskier galaxies also have
slightly  lower
$n$ values, as we would expect, with the median $n$ for galaxies 
with D/T $> 0.5$ being around $2.8 \pm 0.2$ compared to $3.2 \pm 0.05$
for galaxies which are purely ellipticals ( D/T$=0 < 0.1$) The older galaxies
also have slightly higher $n$ values, and the dependence on $n$ on 
$V_{\rm circ}$ is basically flat.

We note that Hopkins \etal (2008) have recently shown that
fitting a single
component Sersic profile can mischaracterize the profile shape of post 
starburst ellipticals. As we have seen with the profiles plotted in Section 
3.2, kinematically selected early types can have low Sersic indices.
Hopkins \etal used two component Sersic fits to examine the inner and outer
profiles of simulated merger remnants. A median  Sersic index for the
outer profile found by Hopkins \etal is $n\sim 2-3$ (e.g. see their Figure 15).
Although our mass and spatial resolution is not high enough to permit two 
component fits here, this issue should be borne in mind when using $n$ to 
separate galaxies into morphological classes (e.g., see Section 4.3)

The specific star formation rate, in the top row of  Figure \ref{36pan}
is pretty flat in the median with respect to the other properties. 
The strongest apparent trend is with galaxy age where it can be seen
that both the median sSFR and the 10\%-90\% envelope do drop with increasing
age. The envelopes do tend to widen also at low  $r_{\rm eff}$ 
and $V_{\rm circ}$, indicating that  galaxies are more extreme
with respect to sSFR when they are small.

\begin{figure}
\centerline{
\psfig{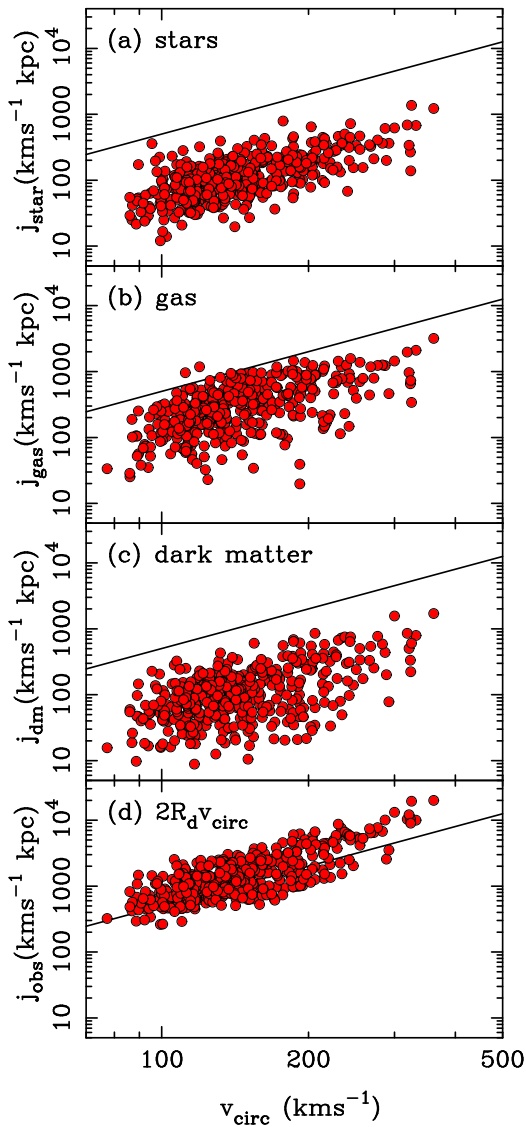}
}
\caption{
Rotation speed vs specific angular momentum $j$ at redshift $z=1$ for
late type galaxies in the {\it BHCosmo} simulation (points). Only galaxies
with a D/T ratio $>0.2$ are shown, the same threshold used to define
late types in the rest of the paper. From top to bottom are 
results for the stellar component,
the gas and the dark matter. The bottom plot shows an estimate
of the specific angular momentum that is not computed directly from
the angular momentum of the particles but instead by multiplying the
disk scale radius by twice the circular velocity. This is the estimator
used for observational data (e.g., Courteau 1997)
and is valid for an exponential disk. The
solid line in each panel is taken from Abadi \etal (2003a), and
describes well the median relation seen in the observational
($z=0$) data compilation of Navarro (1998). In order to compare directly
to Abadi \etal (2003a) we have rescaled using $h=0.65$ when computing the
values of $j$.
\label{jvc}
}
\end{figure}

Looking at Figure \ref{36pan} overall, we can see that there
are many instances of the expected trends between properties which 
we know from observations (e.g., Blanton \etal 2003). For example galaxies are
usually classified into ``early'' and ``late'' types where one can
broadly say that early types are larger, have older stellar populations
and larger bulges than late types. Of the 15 panels in  Figure \ref{36pan}
which carry independent information, 10 of them fit into this simple
picture (for example, galaxies seen kinematically to have larger
disks do have smaller Sersic indices), 1 runs counter to it (we find
that older
galaxies are smaller, as measured by  $r_{\rm eff}$) and 4 are neutral
(sSFR exhibits little median variation with other properties). 
It is beyond the scope of this paper to examine any of these relationships
in detail, but it is apparent that simulations
are nearing the point where in future work it will be profitable to
do so and make more direct comparisons with observational data.

\subsection{Angular momentum}

Simulations of disk galaxy formation have historically had problems
with excessive angular momentum loss, leading to stellar components
with bulges too large compared to observations 
(Mayer \etal 2008 for a review). These include for example
the zoomed simulations of Sommer-Larsen \etal (2003), Abadi \etal 
(2003ab), Robertson
\etal (2004).
 Robertson \etal (2004)  and Abadi \etal (2003ab)
pointed out that exponential 
galactic disks could be formed
which were of the correct size, but which were deficient
in specific angular momentum. Governato \etal (2007)
and Ceverino \& Klypin (2007) have most recently
been able to form disks which were closer to being realistic, through
a combination of high resolution and modelling of star formation and
feedback. All of these simulations were zoomed runs, however, and it is
not clear to what extent these results are dependent on the histories
of the individual halos selected for resimulation.

A clear way to describe the extent of any potential angular momentum
problem is to plot the specific angular momentum content of
galaxies as a function of their circular velocities. This has been done
by for example Navarro \& Steinmetz (2000) and
Abadi \etal (2003a), comparing the 
simulation results against the observational compilation of Navarro (1998).
 Observations of the specific stellar angular momentum
content of disk galaxies were made using an 
estimator which combines the observed circular velocity and
disk scale length (Courteau  1997), so that 
$j_{obs}= 2 R_{\rm eff} V_{\rm circ}$.  For the simulations, the circular
velocity is usually computed using Equation 2, but the angular momentum is
summed directly from the star particles.

In Figure \ref{jvc} we show results from the {\it BHCosmo} simulation
at $z=1$. It is important to note that the simulated galaxies we
are plotting are from $z=1$ whereas the observations (Mathewson
\etal 1992, Courteau  1997)
were made at $z\sim0$. Based on the lack of evolution in angular momentum
vs $V_{\rm circ}$ seen in 
the D4 run (which was run to $z=0$) we expect this to make little
difference to the comparison. The straight line in Figure \ref{jvc}
is taken from Abadi \etal (2003a). It describes the mean of the
observational data well. For the simulation points, because we are comparing
to observational data for disk galaxies, we apply a threshhold on the
D/T ratio before plotting points in Figure \ref{jvc}. This threshold
is at a D/T ratio of 0.2, which selects 40\% of the galaxies to be 
disks at this redshift, $z=1$. Although this cut is essentially arbitrary
(we have a smooth continuum of D/T ratios for galaxies) , we
will see later 
(Section 4.3) that it enables us to reproduce very approximately 
the fraction of morphologically classfied galaxies of  early and late
types as a function of density. For the points in Figure \ref{jvc} we 
measure $j$, the specific angular momentum by summing the total 
angular momentum of the particles in question (we deal with stars,
gas and dark matter separately in the different panels) and dividing by
their summed mass. We use only particles within 2.2 times $R_{\rm eff}$,
in order to be consistent with observations, and include all particles
interior to that radius, including those in the bulge.

We can see in the top panel of Figure \ref{jvc} that the median
$j_{\rm star}$ is a factor of $\sim 8$ lower than the observational value
for disk galaxies. As we have seen in the distribution of D/T ratios,
simulated galactic stars have too little orbital angular momentum. This result
is somewhat better than the lower resolution runs of Navarro \& Steinmetz
(2000), who were a factor of $\sim 30$ below the observations, but the angular 
momentum deficit is still large. Robertson \etal (2004),
with somewhat better resolution in their resimulated galaxies 
(a dark matter particle mass of $3\times10^{6} \msun$, 4 times
smaller than Abadi \etal 2003a)
 formed a
spiral with $j_{\rm star}$ approximately a
factor of 2 below the observational trend.
This galaxy, with a quiet late merger history is similar in its
$j_{\rm star}$ to many of our galaxies. Our results are therefore likely
to be similar to theirs, except that we have access to a whole population
of simulated galaxies.

The median gas specific angular momentum in  Figure \ref{jvc} is 
closer to the observational line than the stars. This was also seen 
by e.g., Abadi \etal (2003a) and Robertson \etal (2004), who find that as gas
cools and stars form, angular momentum is lost to the galactic halo.
For the dark matter, the $j_{\rm dm}$ values are closer to those seen
in the stars, (as also seen by  Abadi \etal 2003a), although the scatter
in $j_{\rm dm}$ for a given $V_{\rm circ}$ is larger.

Although the angular momentum content of the disks is low, we find that the
size of the disks is actually reasonable compared to observations. To
show this, we plot in the bottom panel of  Figure \ref{jvc} the observational
estimator for the specific angular momentum of stars, $j_{\rm obs}$.
In this case, we find that the median is compatible with the observational
line. Therefore, as seen by Roberton \etal (2005) in a single galaxy, the 
disks are of reasonable scale, but they have a bulge component which is
too large, so that too much of the galaxy is supported by non-circular 
motions.

\begin{figure}
\centerline{
\psfig{file=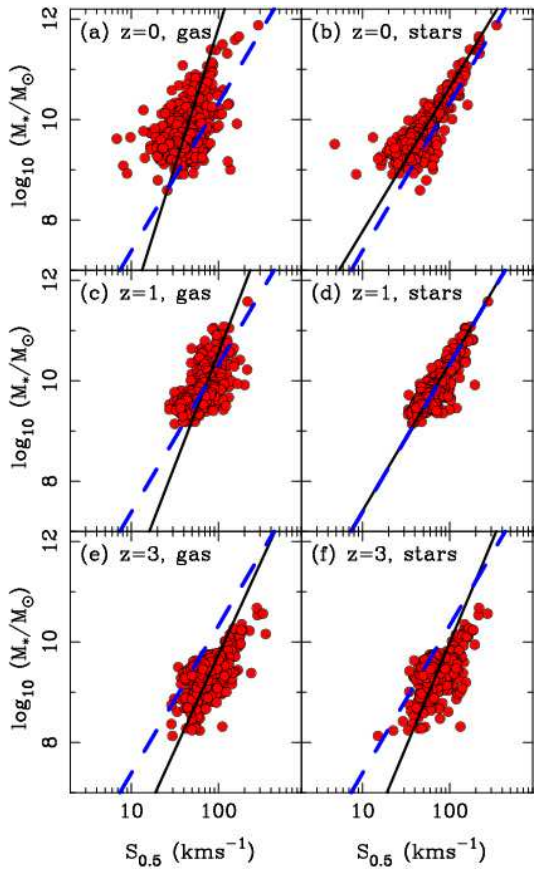,angle=-90.,width=7.0truecm}
}
\caption{
The stellar mass Tully-Fisher relation for late type galaxies (selected
using D/T $>0.2$) in the {\it BHCosmo} simulation (at redshift
$z=1$ and $z=3$) and the D4 simulation ($z=0$). The $x$-axis is
the statistic $S_{0.5}$ from Kassin \etal (2007), which
is a kinematic estimator of support from disordered and ordered motion
(see equation 3).  $S_{0.5}$ from the simulations
was computed either using
gas particle velocities (left panels) or star particle velocities (right
panels). The solid line shows a best fit power law relation to the
simulation data in each panel. The dashed line is the best fit power 
relation from Kassin \etal (2007) to their observational data. This
observational fit
is the average relation measured from 4 bins varying in redshift from 
$z=0.1-1.2$. Kassin \etal (2007) find no statistically significant
variation with redshift.
\label{smtf}
}
\end{figure}

By looking at the angular momentum content of $\sim 500$ disk galaxies, we
see that the problem seen in resimulations of a small number of 
galaxies persists. The addition of black hole
accretion and feedback to our runs
compared to those mentioned above does not noticeably impact the
excess population of non-rotationally supported stars. 
This is compatible with the recent work of Okamoto \etal (2007) who resimulate
disk galaxies with and without black hole feedback.

\subsection{Stellar mass Tully-Fisher relation}

The Tully-Fisher (1977) relation (hereafter TFR)
 between the luminosities of galaxies
and their rotation velocities has long been used as a test 
of galaxy formation models (e.g., Steinmetz \& Navarro 1999). 
For example, the amplitude and
zero point of the TFR seen in observations 
at $z\sim0$ is still  difficult to reproduce
 in high resolution simulations of disk 
galaxies (Portinari \& Sommer-Larsen 2007).
The stellar mass-rotation velocity version of the TFR
is easier to compute theoretically. It has recently been measured 
out to $z=1.2$ using
observational data from the AEGIS and DEEP2 surveys (Kassin \etal 2007).
In the usual M$_{\*}$-V$_{\rm rot}$
version of the TFR, where  M$_{\*}$ is the galaxy 
stellar mass and V$_{\rm rot}$
there is a strong dependence of scatter in the relation on 
galaxy morphology (disturbed, close pairs etc..). 
Kassin \etal (2007) however
find a modification of the TFR which works well with minimal
morphological pruning and holds over the entire
redshift range $z=0.1-1.2$. 

The relation computed by Kassin \etal is between M$_{\*}$
and $S_{0.5}$, where the latter is a kinematic estimator which 
combines dynamical support from ordered motion with that from disordered
motion (see Weiner \etal 2006). The quantity  $S_{0.5}$ is
computed using 
\begin{equation}
S^{2}_{0.5}= 0.5 {\rm V_{rot}}^{2} + \sigma^{2}_{g},
\end{equation}
where  $V_{rot}$ is the rotation velocity of galaxy on the flat
part of the rotation curve,
 $\sigma_{g}$ is a gas 
velocity dispersion, both being measured from a two-dimensional fit to
emission line in the 
galaxy spectra.  Kassin \etal find the 
observed M$_{\*}-S_{0.5}$ relation
to be fairly tight, with an rms intrinsic scatter in $S_{0.5}$
of 0.10 dex. The slope and amplitude of a log-log fit to the points,
$\log_{10} S_{0.5}=a+b \log_{10} M_{\*}/10^{10}\msun$ is $a=0.34 \pm 0.05$ and
$b=1.89 \pm 0.03$ averaged over 4 bins in redshift ranging 
from $z=0.1-1.2$. The fits are not significantly different for all redshift
bins.

We compute  M$_{\*}$ and $S_{0.5}$  values for disk galaxies in our
simulations. Disk galaxies are again taken to be those for 
which D/T $> 0.2$.
 The M$_{\*}$ values are straightforward: we 
 use the total mass in star particles in each galaxy's subhalo. For
 $S_{0.5}$ we have a choice, whether to use a quantity measured from 
particle velocities in the simulation, or whether to use the mass within a
certain radius to compute a value of  $V_{rot}$, assuming rotational
support. In order to include the effect of non-rotational motions, we do the 
former. For $V_{rot}$, we compute the mean rotational velocity 
of particles in a radial
bin of width $r_{\rm eff}$ centered on a radius $2.2 r_{\rm eff}$
from the center of each galaxy, and we compute  $\sigma_{g}$ from
the scatter of particle velocities about  $V_{rot}$. Using these
quantities, we estimate $S_{0.5}$ for the simulated galaxies. For 
each galaxy, we compute two versions of  $S_{0.5}$, one using gas 
particle information, which more closely corresponds to the
observational information, and another using the star particles.

The results are shown in Figure \ref{smtf}, where we show $S_{0.5}$
computed from both gas and stars for 3 different redshifts. For $z=1$
and $z=3$,  we use the {\it BHCosmo} simulation, whereas at $z=0$
we use the D4 run. We also show the power law relation which is
the average one for the data of Kassin \etal (2007) over $z=0.1-1.2$.
For each panel, we also compute the best fit power law relation from our
simulated data and show it as a solid line. In computing the fit, we
follow  Kassin \etal and assign the intrinsic scatter (in this case
we assume all the scatter is intrinsic) to the $S_{0.5}$ coordinate.

We notice from Figure \ref{smtf} that the $S_{0.5}$ values computed from the
gas have a larger scatter about the mean relation than the stars.
The  rms scatter in  $S_{0.5}$  ranges from a maximum of 0.16 dex
for the gas at $z=0$ to a minimum of 0.08 for the stars at $z=1$. This
is not to dissimilar from the observational data, which has $S_{0.5}$
instrinsic scatter rangin from 0.08 to 0.12 dex over 4 redshift bins.
As with the observations, we find no systematic change in rms
scatter with redshift. It can also be seen that fits to
 the stellar simulated $S_{0.5}$ values are consistent with
the observed fit. The largest discrepancy is at $z=3$, a higher redshift
than yet reached with the observational data, for which the slope
is $b=0.24 \pm 0.02$, less than $2\sigma$ different. The  $S_{0.5}$ values 
computed from the gas particles do appear to have a systematically
slightly steeper slope than the observations 
(we find $b=0.22 \pm 0.03$ averaged over all 3 redshifts).
but this is less significant than the $\sim 4 \sigma$ discrepancy
in the intercept, $a$ which is seen at $z=0$.

Overall, we therefore find that the stellar mass TFR relation (between
M$_{\*}$ and $S_{0.5}$) 
measured in the simulations does not change in slope with redshift (between
$z=0$ and $z=3$) and is consistent with the slope seen in the observational
data of Kassin \etal (2007). The zero point measured from the gas
properties in the simulation is somewhat noisy, but not systematically
different when averaged over all redshifts.

\subsection{Orientation}

The relationship between a galaxy's morphology and angular momentum can
give important clues to the processes that occurred during its formation.
 A way to measure this quantitively is to measure the alignment
between the first-ranked principal axis of intertia of a galaxy and its angular
momentum axis. We have carried this out for all galaxies in the
{\it BHCosmo} simulation, again at $z=1$, including the low D/T
ratio galaxies.

\begin{figure}
\centerline{
\psfig{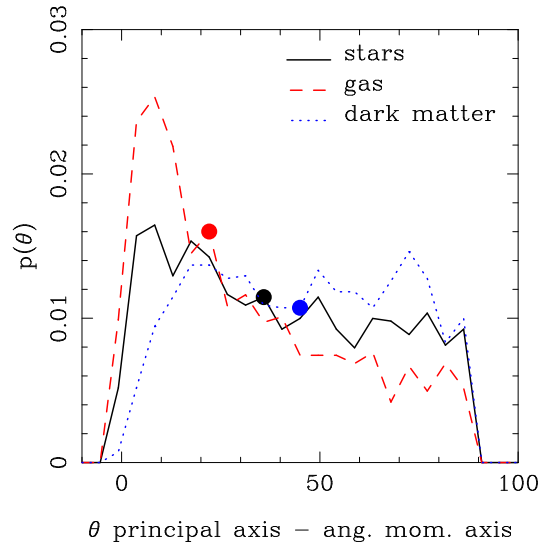}
}
\caption{
Histogram of the angle (in degrees) between the principal 
axis measured from the 
inertia tensor and the angular momentum axis, for galaxies in the
{\it BHCosmo} simulation at redshift $z=1$. We show results for the stars,
gas and dark matter distributions, but the angle in each case is measured
from the angular momentum axis computed from the stars. The solid points
show the median angles (25.7$^{\circ}$ for gas, 39.0$^{\circ}$ for 
stars, and 48.3$^{\circ}$ for dark matter).
\label{principal}
}
\end{figure}

We compute the inertia tensor for each galaxy
 and diagonalize it, to find the
principal axis with the largest moment
of inertia. We do this separately for each component
of the galaxy, using every particle in the subgroup.
From the scalar product of this axis with the
angular momentum axis, we find an alignment angle, $\theta$.  The 
pdf of $\theta$ values is shown in Figure \ref{principal}. We find that
the gas component of galaxies is most strongly aligned,
with a pronounced peak close to zero degrees, likely to be coming 
from the gas disks seen in many galaxies (for example, even the
early type galaxy in the top panel of Figure \ref{plotellip} had
a pronounced gas disk. The median value for the gas is $\theta=25.7^{\circ}$,
significantly more aligned than
the stars ($\theta=39.0^{\circ}$). The stellar
alignment angles do have a low $\theta$ peak, however, indicating that 
rotationally supported stellar disks are still common. By looking at the
orientations of galaxy shapes in Figure \ref{plotellip} (where the
projections are defined by the A.M. axis) we can see that the misaligned
galaxies are indeed likely to be mostly early types, as we would expect.
Finally, in Figure \ref{principal} we can see that the dark matter halos are
also noticeably flattened perpendicular to the rotation axis (the same
flattening direction as the gas and stars),
 with a median alignment angle
of $\theta=48.3^{\circ}$. The expectation value for the median 
of a  random distribution is $\theta=60.4^{\circ}$, with a Poisson error of
$1.8 ^{\circ}$. As the dark matter is dissipationless, this flattening
could be partly due to adiabatic contraction in response to the 
rotationally flatted gas and stellar component. A strong effect was also 
seen by Bailin \& Steinmetz (2005) in purely dissipationless simulations.

\begin{figure}
\centerline{
\psfig{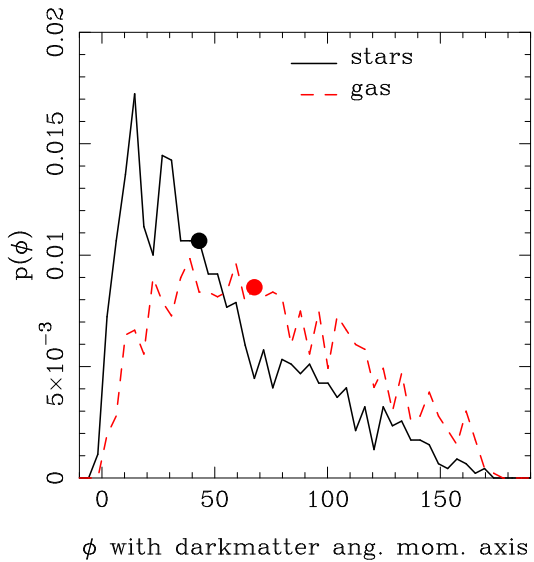}
}
\caption{
The alignment angle, $\phi$ (in degrees) between the star and gas
angular momentum axes and the dark matter angular momentum axis,
for galaxies in the {\it BHCosmo} simulation at redshift $z=1$.
The solid points are the medians of the distributions (43.5$^{\circ}$
for stars and $69.6^{\circ}$ for gas).
\label{amaxis}
}
\end{figure}

Looking at the alignments of the angular momentum
vectors of the different components themselves
can also give us some clues to how the rotational support arises. In figure
\ref{amaxis} we plot the distribution of  alignment angles $\phi$ 
of the star and gas angular momentum axes with the dark matter
angular momentum axis for each galaxy.  We find that there is a strong
alignment of the stellar A.M. with the dark matter A.M. (median 
angle $\phi=43.5^{\circ}$), and a somewhat weaker alignment of the gas
with the dark matter (median $\phi=69.6^{\circ}$). This is quite suprising
given the alignment of the principal axes of the gas seen in Figure 
\ref{principal}. We have also computed (not plotted) the alignment of 
gas and star angular momentum vectors. We find a median angle in this case
of $\phi=61.6^{\circ}$, closer to the alignment of the gas and dark matter. 
We can therefore see that the stellar angular momenta appear to have
become correlated with that of the dark matter {\it after} the 
stars have formed from the gas. This coupling through collisionless 
dynamics could perhaps be a numerical effect, due to two body interactions.
To provide additional information,
 we have investigated our lower resolution simulation,
 the D4 and found
that the rank order of the median alignment angles are the same as
for the {\it BHCosmo} run, i.e. star-dark, star-gas, gas-dark, in order
of best to worst alignment. The median angles of the latter two are
$\sim 10^{\circ}$ worse,  but for  star-dark, the angle is $\sim 20^{\circ}$
worse, suggesting that the stellar angular momentum is affected by the
most by lack of resolution.

We note that in the simulations of Van den Bosch \etal (2002), a
much stronger alignment was seen between the angular momentum of the
(adiabatic) gas in halos and the dark matter (median angle  
$\phi=27.1^{\circ}$) than here, so 
that without dissipation, the gas is much more 
strongly coupled in its rotation to the dark matter.

\subsection{Age of disk and bulge stars}

\begin{figure}
\centerline{
\psfig{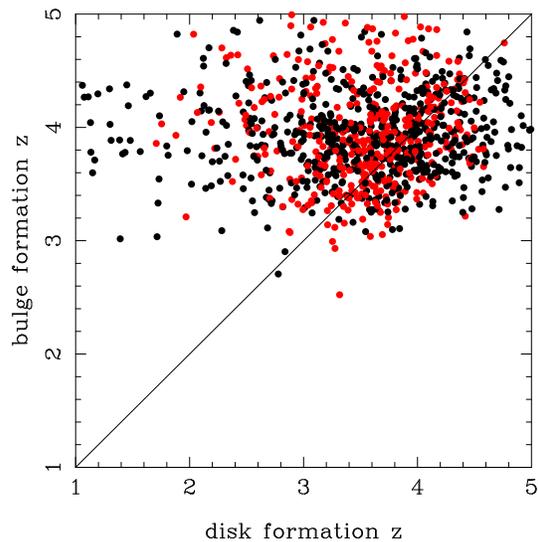}
}
\caption{
The mean age of star particles defined kinematically to be in 
a disk component, versus the age of those in the bulge, for galaxies in 
the {\it BHCosmo} simulation at redshift $z=1$. We show results for 
late type galaxies (D/T $>0.2$) as red points and early types as black 
points. 
\label{dbage}
}
\end{figure}

We have seen in Figure \ref{36pan} that galaxy ages do have some dependence
on other properties, including size and D/T ratio. We have decomposed
the galaxies kinematically into bulges and disks (Section 3.1), and 
so it is of interest to see whether the ages of the disk and bulge
stars fit into this general picture. In Figure \ref{dbage}, we have
plotted the mean redshift of formation of stars in the disks and bulge
components of the galaxies in the {\it BHcosmo} run at $z=1$. We show the
galaxies with D/T ratios greater than 0.2 (our fiducial threshold between
kinematically defined early and late types) as red points.

We can see that the cloud of points extends down to low redshifts,
indicating that the disk component of galaxies does indeed 
form significantly later than the bulge. For example, there are essentially
no bulges with a mean star formation redshift below $z=3$, but 
disks continue to form all the way down to $z=1$, the lowest redshift
possible. The disks below $z\sim2$ are all disk components of
bulge dominated galaxies, so that they only represent a small fraction
of the galaxy. Interestingly, we also see that the disks with the
highest formation ages $z \simgt 4.5$ also belong to early-type galaxies.

\section{Relation between environment and galaxy properties}

In the previous section we have seen that individual galaxy properties
in the simulations are by and large what we would expect given
 a mixture of early and late types. We will now investigate how the
immediate environment has affected these properties.

\subsection{Measures of environment}

We use four different definitions of the local environment of a
galaxy, in order to span
a range of parameters from those close to what can 
be measured observationally to measures related
directly to the dark matter
distribution. The first measure is the density computed from the 
distance to the tenth nearest galaxy. For this we use positions for 
galaxies above our 5000 particle threshold, and compute a local density
for each galaxy, dividing the density of the sphere enclosing 10 neighbours
by the mean density of galaxies in the simulation. We label this 
quantity on the plots as $\rho/\left< \rho \right>$.
 The density computed from the
tenth nearest galaxy is often used in observational measurements,
such as in the morphology-density relation of Dressler (1980), 
and subsequently  by Dressler \etal (1997), Smith \etal (2005),
Carpak \etal (2007) and others.
In these cases, the measurement is made in
2 dimensions, from the projected positions of galaxies, but the 
two quantities will be closely related. We explore this measure more closely 
in the next section.

A related measure is $\rho_{13}$, which is the density computed from 
a sphere centered on the galaxy which contains $10^{13} \msun$
of dark matter. In practice, this will be very similar to the first 
measure, based on galaxies, but will be less noisy. It will also
be more directly related to the gravitational   influence of nearby structures,
being based on mass density and not number density. We note that a very 
similar measure of environment, computed from the radius
of a sphere which contains $10^{13} \msun$ of all types of matter
was used by Colberg \& Di Matteo (2008) in their study of the environmental
dependence of black hole properties in the same simulation (the
{\it BHCosmo} run) which we use here.

The galaxies are members of friends-of-friends groups of particles, and
so we can also categorise their environment using these grops.
We  use the mass of the FOF 
group as one means of quantifying their environment, and the other
the distance of the galaxy from the FOF group center of mass, in units
of the virial radius, $r_{200}$. We compute $r_{200}$ by measuring the
radius of a sphere which contains 200 times the mean density.

\subsection{Galaxy properties and environment at z=1}

\begin{figure*}
\centerline{
\psfig{file=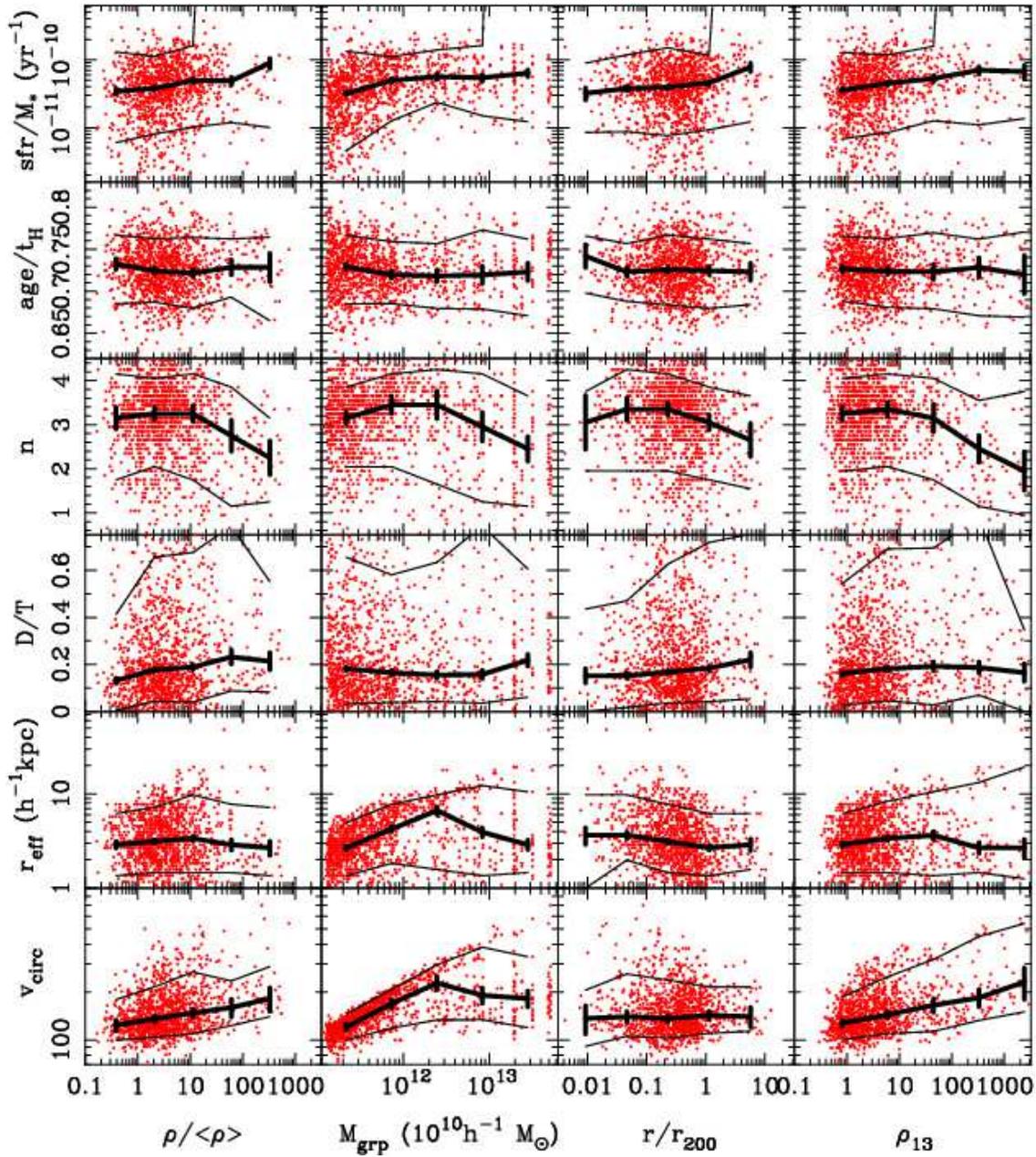,angle=-90.,width=15.0truecm}
}
\caption{
Galaxy properties versus environment at $z=1$ in the
{\it BHCosmo} simulation. The measures of environment are, from left
to right, galaxy density (measured from the
distance to the 10th nearest neighbour galaxy) in units of the mean,
mass of the host friends-of-friends group, distance from the center
of mass of the friends-of-friends group, in units of the
virial radius, $r_{200}$, and dark matter density computed from the radius
which encloses $10^{13} \msun$ of dark matter. In each panel,
results for individual galaxies are shown as points, and the median is
shown as a thick solid line with Poisson error bars. The 90th and
10th percentiles of the distribution in each bin are shown as thin 
solid lines.
\label{z1rhomorph}
}
\end{figure*}

In Figure \ref{z1rhomorph}, we use the same galaxy parameters which 
were shown in Figure \ref{36pan}, but this time instead of plotting them
against each other, we plot them against our four measures of
local environment. We again show the situation at $z=1$, for
the {\it BHcosmo} simulation.
Before proceeding with our analysis we note that at $z=1$
there are no rich clusters in the relatively small simulation volume.
(the largest FOF group has a mass of $6\times10^{13} \msun$).
Below we will see that this will affect our ability to compare
to observational data for the high density end of environments at this
redshift. In interpreting  Figure \ref{z1rhomorph} we should therefore
bear in mind that the dense environments in the plot are
groups rather than clusters. In order to look at galaxies in clusters
at high resolution, a resimulation approach should be used (see for example
Saro \etal 2006.)

Looking at the figure, we can 
immediately see that the results are a mixture of relationships
which we would have expected and some which are surprising. The
 values of $V_{\rm circ}$ for galaxies increase steadily
with density, both $\rho/\left< \rho \right>$ 
and $\rho_{13}$, the median rising
by approximately a factor of 2 over 3 orders of magnitude in density.
For  $V_{\rm circ}$ vs ${\rm M_{grp}}$, we see that there is an upper
envelope of large galaxies in each group, but also that when 
${\rm M_{grp}}$ reaches $2\times 10^{12}  \msun$, there are enough satellite
galaxies to pull the median  $V_{\rm circ}$ value down.
In terms of $r/r_{200}$, we can see that the galaxies with the 
very highest $V_{\rm circ}$ values are indeed close to the center
of mass of the FOF groups. There is a large cloud of points,
between $0.1$ and $\sim 1.0$ virial radii from the center, indicating where
most of the galaxies are found. The number density of points decreases
rapidly beyond  $r/r_{200}$, although a few galaxies are found at 
distances of up to 10 virial radii. The median of  $V_{\rm circ}$ values
does not vary with $r/r_{200}$.

The variation of $r_{\rm eff}$ with density is similar to $V_{\rm circ}$,
except that the effect of satellite galaxies appears to be more
pronounced. Also, even the median $r_{\rm eff}$ value changes with $r/r_{200}$.

Moving on to the D/T ratio in the third row from bottom of 
Figure \ref{z1rhomorph}, we find more complex dependencies on density. 
There is a moderate rise in the median D/T ratio,
from $0.13$ to $0.2$ as $\rho/\langle \rho \rangle>$
changes by 3 orders of magnitude. It therefore seems as though there
are many small bulge-heavy galaxies at low densities.  This behaviour
is also repeated but more weakly
(only significant
at the $\sim 1\sigma$ level) when looking at $\rho_{13}$. If we
look at the relationship between  D/T and $r/r_{200}$ we can see that
the median D/T rises towards the edges of groups,
significant at the $3-4 \sigma$ level, and the 90th percentile
line increases dramatically. There are therefore not many disk dominated
galaxies in the centers of groups. This seems to be opposite to the
trend seen in the variation of D/T  with density. However,  it is
possible that the D/T values do go down with very high densities, but that 
this is not easy to see with the relatively noisy density estimators that 
we have (the 90th percentile of  D/T does appear
to decrease for large $\rho_{13}$.) When comparing   $r/r_{200}$ and
 $\rho$ values we  should bear in mind that when large galaxies are
present, the group center of mass better correlates with the centers
of large galaxies, and hence the density. This can help explain
some of the apparent differences in behaviours with  $r/r_{200}$ 
and $\rho$.

The variation of Sersic index, $n$ with density is basically 
consistent with flat from
low to moderate density, followed by a decline in the median value, indicating
fewer centrally concentrated early types at high density. This
is somewhat consistent with the trend seen in D/T ratios, although in that
case more variations were seen at low density than high. For $n$, the
errors are large, but it does seem from the 90th and 10th percentiles
that late types preferentially prefer the outskirts of groups as well.

The mean stellar age of galaxies does not appear to vary much with density
at this redshift.
There is perhaps even a small hint that galaxies in low density
regions are slightly older, something which is puzzling but consistent
with the slight preponderance of bulge-dominated systems there. 

The specific star formation rate of galaxies in
our simulations rises slightly with 
density, but also on the outskirts of groups. While the trend seen
in observations at $z=1$ by Cooper \etal (2008) has the same
sign as that seen at low redshifts (e.g., Gomez \etal 2003),
Cooper \etal find that the total SFR has an inverted relation 
with density. The  behaviour and comparison 
with observations, particularily of the total SFR
 is investigated fully in this
{\it BHCosmo} simulation by Colberg \& Di Matteo (2008). For now,
we note that the specific SFR trend we plot
is roughly consistent with what has been seen
with the D/T ratio and Sersic index, i.e., a reversal, or at least flattening
of the trends seen in galaxy morphology with environment
at redshift $z=0$.

Some of the trends of galaxy properties
with density do have some dependence
  on galaxy mass/particle number. For example
when we limit ourselves to plotting galaxies containing 
25000 particles instead of 5000 (although we still use all galaxies
to compute  $\rho/ \left<\rho \right>$), 
we find that the median D/T value for the highest $\rho/ \left<\rho \right>$
bin drops by 2 $\sigma$ to D/T=0.1. The previous bins are essentially
unchanged. It is therefore possible that the efficiency
of destruction of spirals
in the highest density regions depends on  particle number.  
In this test, the qualitative behaviour
of SFR is unchanged, but the highest density bin does have slightly older
galaxies.

Our overall conclusion from Figure \ref{z1rhomorph} is that at $z=1$ the
relationships between galaxy properties 
 and density in our simulations are complex. The 
preponderance of late types in low density regions seen at $z=0$ seems
to be reversed, with higher star formation rates in high densities. At the
very highest densities, there are hints that  the disk fraction may decrease,
but this is dependent on resolution/particle number, with the main
uncertainty that our simulation box, at $33.75 \hmpc$ is too small
to contain any clusters at $z=1$. Cooper \etal (2007) have investigated the
observed 
relationship between galaxy color and local density from redshifts $z=0.4$
to $z=1.35$, finding that the  trend of red sequence
galaxies to favour overdense regions evolves strongly over that redshift
range. The locations of red galaxies $z\sim1$ are however still seen to 
correlate with high densities. Population synthesis models will be needed
to compare colors of our simulated galaxies with this information. For now,
we restrict ourselves to the  morphological information we have computed,
although as we shall see in the next section, the morphology-density relation
in the simulation also appears to evolve somewhat more weakly than in the
observations.

\subsection{Morphology-density relation}

\begin{figure}
\centerline{
\psfig{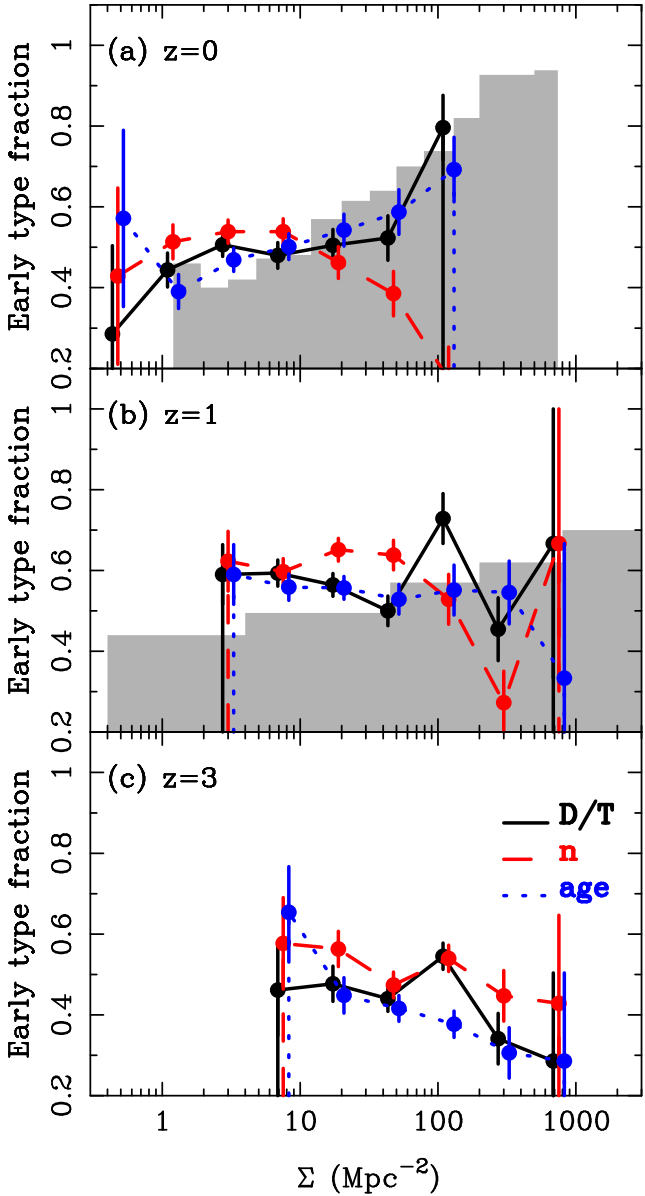}
}
\caption{
The morphology-density relation at 3 redshifts. We show as different
lines results for 
galaxies types selected on the basis of kinematic D/T ratio, 
Sersic index, and mean stellar age (see text).
The $x-$axis, $\Sigma$ is the projected galaxy density.
 Panels for $z=1$ and
$z=3$ are for galaxies from the {\it BHcosmo} simulation, and 
$z=0$ galaxies are taken from the D4 run. As shaded histograms in
the top 2 panels we show the observational results of Dressler (1980,1997)
at $z\sim0$ and Smith \etal (2005) at $z=1$.
\label{morphden}
}
\end{figure}

The tendency of dense (galaxy cluster) environments in the low redshift 
Universe to contain a higher fraction of early types than the field has 
long been known (Dressler 1980). Recently it has become possible to extend
this work to higher redshifts, $z=0.5-1$ and beyond.
Here we compare simulation results for this morphology-density relation
to the observational data of Smith \etal (2005),
which, along with data from $z=0.75-1.27$ includes a compilation
at lower and intermediate redshifts from 
Dressler (1980, 1997), and Treu \etal (2003). 

 The galaxies in the 
Smith \etal sample and compilation
 have limiting V magnitude of -20.2 at $z=0$ (-21.2 at z=1).
 Their space density (using the luminosity function of
 Brown \etal 2001) is 
$5.12 \times 10^{-2}$ $(\hmpc)^{-3}$   at z=0.
The observational fraction of early-type galaxies is plotted as a
function of projected galaxy density at 2 different redshifts in the
top 2 panels of Figure \ref{morphden}. The projected galaxy density is the
density enclosed within a rectangular region extending to the
10th nearest galaxy. In order to include redshift information in
the $z=1$ data, Smith \etal
include galaxies in the density estimate  in slices of redshift width
 $\Delta z=0.1$. They also subtract a background galaxy density
from the density estimate and also multiply by a density correction factor
of 2. After this procedure, their results are
 that the correlation
between early type fraction and density is still present at $z=1$, but
significantly weaker than at $z=0$  (and also $z=0.5$, which lies in 
between, and is not plotted here).

In order to make predictions from the simulations, we also compute a projected
galaxy density measured from the area delimited by the 10th nearest neighbor.
It is not possible to replicate entirely the relatively involved procedure
employed by Smith \etal, but we make sure that we are centered on
the  same mean
projected galaxy density, and make sure that our results are robust by also
trying a completely different density estimator, 
as explained below.

We project the whole simulation box, using information from the 3 different
orthogonal projections in our data sample. This results 
in a lower mean projected galaxy
density than Smith \etal, who at $z=1$ 
have a mean area density  of galaxies
of  $5.25$ per Mpc$^2$. Our galaxy density is 2.0 per Mpc$^2$,
so that we correct our projected simulation density values by a constant
factor to bring them into agreement. This is under the assumption
that the overdensity of galaxies relative to the mean is the 
physically relevant quantity. 
We have also tried computing the
morphology density relation using  the 3 dimensional galaxy overdensity from 
the simulations, for comparison. We find that all the qualitative
conclusions we
draw from the comparison with projected observational data also hold,
giving us confidence in the robustness of the results.

In order to find the early type fraction, we compute
three different values using  three
different criteria, kinematic D/T ratio,  Sersic index and age.
For the kinematic selection, we use a cutoff of D/T=0.2 between early
and late types, and  for the Sersic index, a threshold of $n=3.0$,
For the age criterion, we used a  cutoff mean stellar
formation redshift between early and late types
which is different for the 3 redshift bins  ($z=0.0$, 1.0 and 3.0) we use.
In order to set the threshold, we choose values which make the overall
fraction of early and late types the same as for our D/T ratio
criterion. This cutoff stellar formation redshift is $z=2.6, 3.8$ and 4.8
for redshift bins $z=0, 1.0$ and $z=3.0$ respectively.

Our results for the morphology-density relation in the simulations
are shown in Figure \ref{morphden} alongside the observational
results at $z=0$ and $z=1$. We also show the simulation
predictions for redshift $z=3$. We can see that at $z=0$, the early
type fraction increases in high density regions. Although there are
no rich clusters in our relatively small volume, the simulation
curves with early types selected kinetmatically by and
age track reasonably well the observational results up to moderate
galaxy density. The photometric criterion however diverges at the high
end of the density range, with the number of galaxies with $n>3$ falling
dramatically at projected densities $> 20 Mpc^{-2}$. This behaviour was also 
seen in Figure \ref{z1rhomorph}. 

At redshift $z=1$, the morphology-density relation in the simulation
flattens out, with no increase seen in the fraction of early types
at high densities in the middle panel of Figure \ref{morphden}. This
change between $z=0$ and $z=1$ is somewhat more extreme than that which is
seen observationally, although the sign of the change is
the same. Again the selection by Sersic index gives the
most discrepant results. At $z=3$, the simulations predict a complete reversal
of the morphology density relation, with the fraction
of galaxies with younger stellar
populations being highest in the densest regions. It is to be noted that 
because the units of the $x$-axis of Figure \ref{morphden} are in proper
Mpc, the galaxy density of the different redshift samples is
affected by the expansion of the universe.

Because the simulation volume is relatively small, we are restricted to
computing the morphology-density relation for galaxies with a relatively 
small range of stellar masses (there are few extremely massive galaxies
for example, particularly at $z=1$ and above). Recently, it has been found
(F. van den Bosch, private communication, Weinmann \etal 2008)
 that for (satellite) galaxies in a fixed
stellar mass bin, there is no significant morphology dependence on host
halo mass, i.e. observationally the morphology-density relation seems 
to be driven by the
stellar mass-density relation.
The results we find in the simulation for the morphology-density relation
could be therefore be relatively flat at least partly
because of our small range of galaxy masses.
We note that we have accounted for differences between central and
satellite galaxies in our study, whereas Weinmann \etal 
find that their result holds for satellite galaxies only.
 Larger simulations will
be needed to investigate conclusively whether the morphology-density
relation in simulations is flat for bins of galaxy stellar mass.

\begin{figure}
\centerline{
\psfig{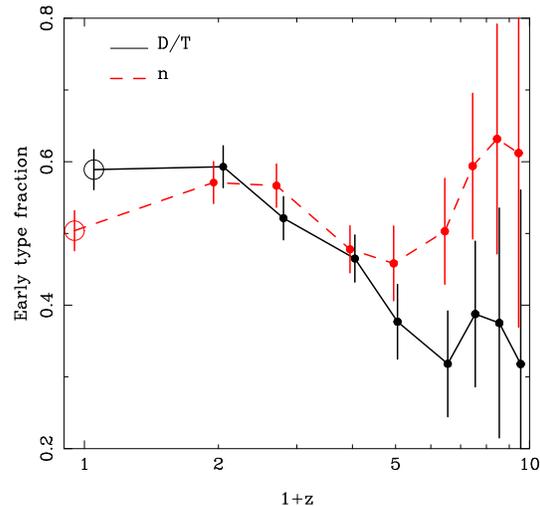}
}
\caption{
The early-type fraction of simulated galaxies as
a function of redshift. Galaxies were classified either on the basis
of kinematic D/T ratio ($>0.2$ for late type) or Sersic index, ($n < 3.0$
for late type), as in Figure \ref{morphden}). The filled points 
(i.e. $z\geq1$) are for the {\it BHCosmo} simulation and the open
points for the D4 run.
\label{z}
}
\end{figure}

We have seen that the morphology-density relation flattens out or inverts
at higher redshifts. However, what is not easily seen in Figure \ref{morphden}
is that the overall fraction of early type galaxies  change 
with redshift as well. We have applied the same thresholds to separate
early and late types (D/T $> 0.2$ and $n < 3.0$ for late types) in outputs
of the {\it BHCosmo} run at redshifts up to $z=9$.  The results are shown 
in Figure \ref{z}, where we can see that the two means of separating 
galaxy types do agree at $z=4$ and below. At high redshifts, although the
errors are large, the correspondence between type chosen through
 D/T and $n$ is not good, with galaxies appearing to  become
steeper of profile but at the same time having smaller 
kinematically defined bulges. Based on the resolution tests 
(Figure \ref{matchmorph}), D/T is likely to be a more robust discriminator
between galaxy types. This conclusion is understandable if we take into
account the fact that single component Sersic profiles are often not
a good characterization of the profile shape of post-starburst ellipticals,
as found by Hopkins \etal (2008) and discussed further in Section 3.4.

 At lower
redshifts, however, the  number of early types does increase significantly.
As described by D/T ratio, the early type fraction increases from 
$\sim 0.35$ at $z\sim9$ to $\sim 0.6$ at $z=1$.
The early types chosen by $n$ also increase below $z=4$
 Following
galaxy properties in detail with redshift
is beyond the scope of this paper, so we
will confine ourselves to noting that the general decrease in the fraction
of disk galaxies as the universe evolves seen here
 is the same qualitative trend seen observationally (e.g., Postman \etal 2005).

\section{Large-scale structure and clustering}

The galaxies in our simulations form at the same time as the cosmic
web of large-scale structure. We have seen from the previous
section how the local environment of galaxies affects their properties.
We will now turn to larger scales, examining the morphology of this structure
as traced by different types of galaxies and also their clustering,
measured using the autocorrelations function. As with the previous section,
 the relatively small simulation volume means that we will not have
instances of truly rare events, rich clusters  of galaxies, and also
the box size will have some effect on the overall amplitude of clustering.
For this reason, we will concentrate on the differences between dark matter
and galaxy clustering, their relative bias, for different galaxy types. This 
should be less sensitive to the small box size than the raw
amplitude of the autocorrelation function (following the 
approach of e.g., Katz \etal 1999).

\subsection{Visual impression}

\begin{figure*}
\centerline{
\psfig{file=slicegals.ps,angle=-90.,width=16.5truecm}
}
\centerline{
\psfig{file=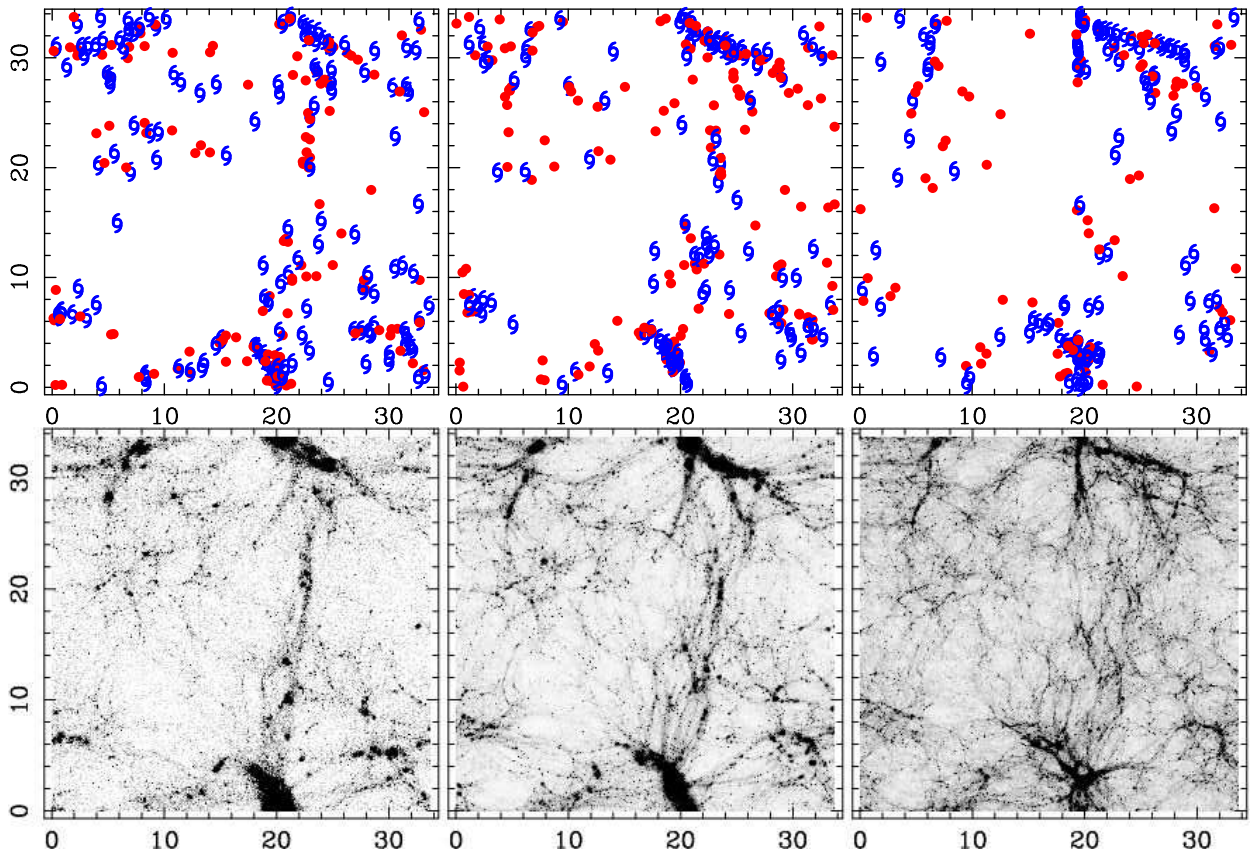,angle=-90.,width=16.5truecm}
}
\caption{Large-scale structure in the galaxy and dark matter distribution.
Top row: slices ($10 \hmpc$ thick) through the simulated galaxy distributions
at 3 redshifts. Late type galaxies (selected so that $D/T >0.2$)
are shown as spiral symbols and early types as points. The leftmost
panel, $z=0$ is for the D4 simulation, and the two right panels ($z=1$,
middle and $z=3$, right) are from the {\it BHCosmo} run. 
Bottom row: a linear greyscale plot of the dark matter distribution
in the same regions of the same simulations shown in the top row.
\label{slicegals}
}
\end{figure*}

In Figure \ref{slicegals} we show $10 \hmpc$ thick slices through the galaxy
and dark matter density fields, at 3 different redshifts. We have
used our usual  D/T theshold of 0.2 to classify galaxies into 
early and late types, which are plotted using different symbols. The same
5000 particle lower limit for subhalos is again used at all redshifts, except
for the leftmost panel, which is taken from the D4 run. In this case, we
use the equivalent particle limit, scaled from the {\it BHCosmo} run
using the ratio of mass resolutions.
 
Looking at the three different redshift
panels, the overall framework of structures looks quite similar
at different times. In the dark matter plots, 
the dense regions at $z=3$, are quite web-like, made up of many
intersecting filaments, whereas at $z=0$ they have collapsed into 
more compact structures. On larger scales, however, not much appears
to have changed, particularly when looking at the galaxy distribution.

Comparing the distribution of early and late types, we see
some evidence for dense clumps of early types at $z=0$ (for example at the
bottom of the panel). There may also be marginally more late types around
the outskirts of these structures, although the situation
does not look markedly different at higher redshift. At redshifts $1$
and $3$, however there do seem to be many early types which 
are relatively isolated, somewhat more than late types.
More conventionally, the protogroup of the bottom of the panel appears
at redshift $3$ to consist of more late types, and then at $z=0$ it
 collapses into a virialized group of early types. Apart from this,
however the visual impression at $z=1$ and above is that late types 
are marginally more highly clustered in the simulation. We shall see
below that this is borne out by quantitative measurement.

\subsection{The autocorrelation function}

We use the same delimitations between galaxy types as in Figure \ref{morphden}
to investigate the autocorrelation function of early and late types. We 
look at three redshifts, $z=3$, $z=1$ and $z=0$, using the D4 simulation
for the latter.
In addition, we separate the full sample into two sets, by mass, in
order to see  whether the mass dependence has a larger effect on
clustering. In order to compare the results directly with those
of galaxy types, we chose to make the higher mass subsample
in each case have the
same number of galaxies as those with D/T $< 0.2$.

\begin{figure*}
\centerline{
\psfig{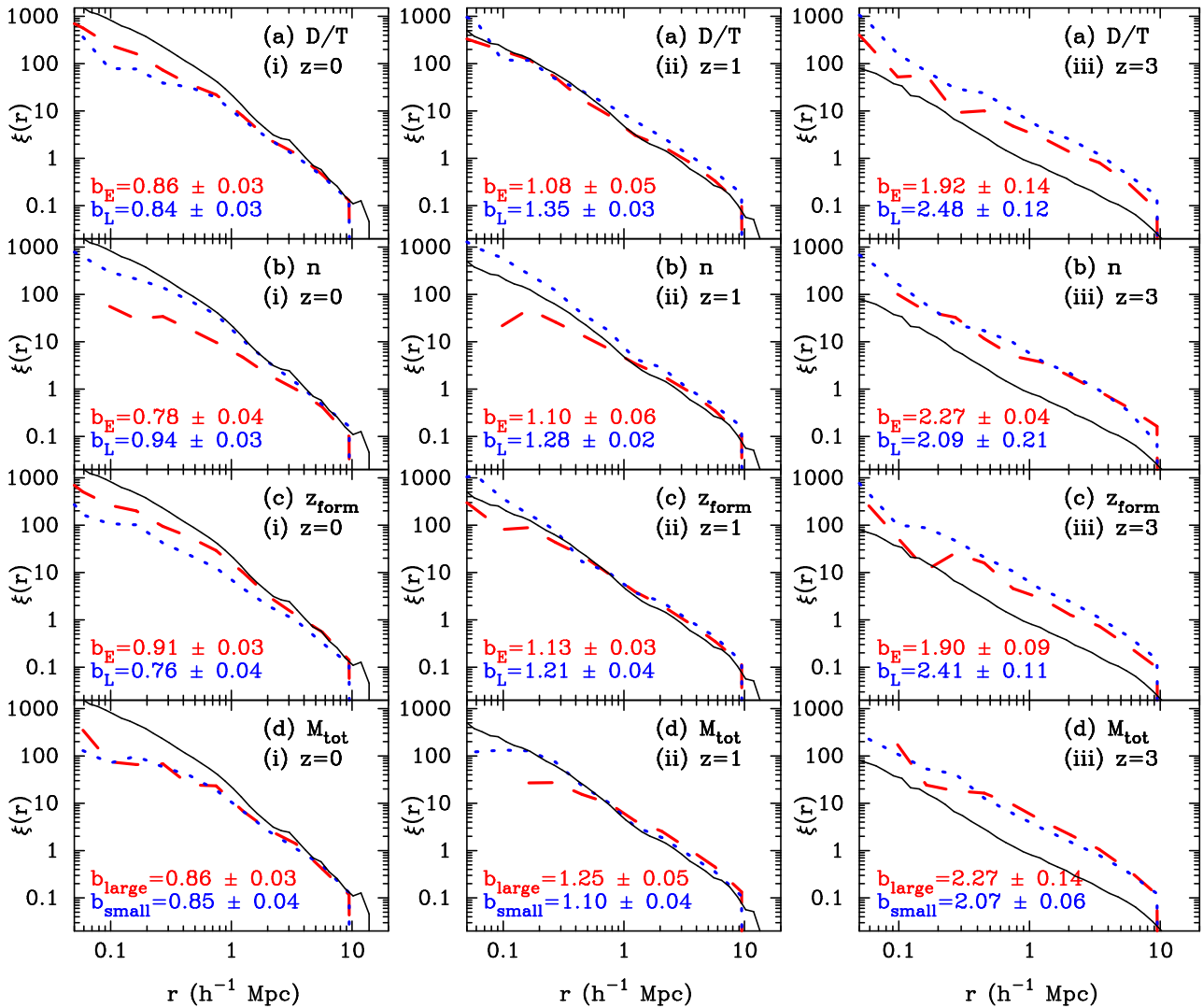}
}
\caption{
The autocorrelation function of simulated galaxies as a function of
type and redshift. The 3 columns show results for $z=0$,
$z=1$ and $z=3$, from left to right, with the $z=0$ results taken
from the D4 simulation and the rest the {\it BHCosmo} run. In each
row, (a)-(c) we show galaxies classfied into
early and late types by D/T ratio, Sersic
index and mean stellar age. In row (d), we show results for
galaxies split into two mass classes (see text). The galaxy correlation
function results are shown as dotted lines for late types (and low mass,
``small'' galaxies in row (d)) and dashed lines for early types. In each 
panel we also plot the autocorrelation function of the dark matter as a thin
solid line. The bias factor between the dark matter and the galaxies is
given in each panel, computed from the average of the bias factor in the
5 bins with  $10 \hmpc > r> 2 \hmpc$. The quoted 1 $\sigma$ errors are 
derived from the standard deviation of results for the 5 bins.
\label{xi}
}
\end{figure*}

We compute $\xi(r)$ in real space, making use of the periodic
boundary conditions of the simulation. The results are shown in 
Figure \ref{xi}. Also given in each panel are values for the large-scale
bias between galaxies and dark matter.
For the early types it is computed using  $b_{\rm E}=\sqrt{\xi_{\rm E}/
\xi_{dm}}$ 
and analogously for the late types. We also compute the bias
for the high mass and low mass subsamples.
In each case we give the mean value for the 5 bins between $2 \hmpc$ and 
$10 \hmpc$, and an error which comes from the standard deviation 
of the results for the 5 bins.

On scales $r \simgt 1 \hmpc$ we can see that in 
all panels we have approximately scale
independent bias between galaxies and dark matter. All galaxy subsamples
at $z \ge1$ have a higher amplitude of clustering on these
scales than the dark matter, with the lowest amplitude 
corresponding to $b=1.08\pm0.03$. At redshift $z=3$, this bias can be
as high as $b=2.48 \pm 0.12$, so that the galaxies follow the familiar
``high peaks'' biasing trend where similar mass galaxies at 
higher redshifts are associated with rarer, higher peaks and
are biased more strongly (e.g., Bardeen \etal 1986). The values of bias
decrease until at $z=0$, all galaxies are antibiased with respect to the
dark matter.

The subsamples based on mass also act in the fashion expected from
high peaks theory, with the more massive galaxies being more biased. However,
if we look at the different methods for  classifying
types, we can see
that in many cases the late types are more highly biased than the early types.
At $z \ge1$, no matter how galaxies are classified, the late types
are more highly biased.
The largest difference is for the kinematic
classification, at $z=1$ for which high
 D/T galaxies are $25\%$ more biased. Even for galaxies
classified based on mean stellar age we see this trend,
with the youngest galaxies (assumed to be late types) being
$7 \%$ more biased at this redshift.

We therefore find that at high redshifts, the disk galaxies have a higher
correlation function ($b=1.35\pm0.03$ at $z=1$) on large scales
than the high mass galaxy subsample ($b=1.25 \pm 0.05$ at $z=1$),
 even though
they actually have a lower median circular velocity ($v_{\rm circ}=136 \kms$
vs $v_{\rm circ}=177 \kms$).
This is quite interesting, implying that the processeses
which determine whether a galaxy is an early or late type at these
redshifts are even more closely coupled to clustering than 
galaxy mass. 

At redshift zero, the situation reverses itself for the most part,
with early types chosen on the basis of D/T and mean stellar age being 
more clustered than the late types. The difference in the clustering of the
large and small mass subsets also becomes statistically insignificant.

If we now turn to  small scales $r \simlt 1 \hmpc$,  the clustering between
galaxies and mass does become scale-dependent. Selecting
galaxy type using $n$ changes the amplitude the most, with early types
being a factor of $\sim 20$ times more clustered at a separation
of $r=0.1 \hmpc$ at $z\le 1$. This can be compared with the 
D/T plot, for which 
kinematic separation has little effect on $\xi$. The 
difference in the slope of the
stellar profile fitted for galaxies with different $n$ therefore has
effects on much larger scales than individual galaxies. 
Given that the effects of galaxy exclusion are seen in the large mass
subsample at $z=1$ ($\xi_{\rm large}$ drops to zero at radii presumably
close to twice the radius of large galaxies), one might worry that this
is also happening in the $n$ panel.  However, looking at Figure \ref{36pan}
one can see that that galaxies with $n>3$ essentially sample
the entire $V_{\rm circ}$ range, so this is unlikely to be a problem.

Observationally, several recent studies have begun to probe
the details of clustering as a function of galaxy properties at redshift $z=1$.
For example, Coil \etal (2006) have investigated luminosity dependence
of clustering in the DEEP2 redshift survey, finding that more luminous
galaxies are more strongly clustered than fainter ones, with the dependence
of relative bias on luminosity being even stronger at $z=1$ than $z=0$. For a
 concordance cosmology, the bias seen by Coil \etal varies from $b=1.26-1.54$.
The number density of galaxies for the observed sample with the lowest
luminosity threshold is $1.3 \times 10^{-2} (\hmpc)^{-3}$,
within $10\%$  of the space density of our simulation sample
of ``large'' galaxies plotted in panel (d) of Figure \ref{xi}. The observed
 value of the bias from Coil \etal is $b=1.26\pm0.04$, compared
to our simulation value of $b_{large}=1.25\pm0.05$. The
observed  clustering 
amplitude for a given space density therefore appears to be reproduced by
the simulations. We note that in this comparison, the bias measured by
Coil \etal is the linear bias whereas we measure the bias with respect to
the non-linear dark matter. At this relatively high redshift 
and on the scales we measure  we do not expect
there to be a significant difference.
 We leave the investigation of clustering as a function
of luminosity and color and exploration of results for higher redshifts 
(where the simulation bias predictions are much higher) to future work.

Meneux et al (2008) have carried out a detailed study of clustering
of galaxies as a function of stellar mass at $z\sim1$ (actually $z=0.85$
mean redshift) in the VIMOS-VLT
Deep Survey. This can be compared directly to our simulation results. To do
this,  we have split up our $z=1$ sample into the same stellar mass
bins as Meneux \etal. These are $\log (M/M_{\odot})=9.5-10.0$,
$\log (M/M_{\odot})=10.0-10.5$, and
$\log (M/M_{\odot})=10.5-11.0$.
Meneux \etal find a bias of $1.29\pm0.10$, $1.32\pm0.10$ and
$1.62\pm0.18$ for these three bins, respectively. Our results are
$0.86 \pm 0.05$, $1.05 \pm 0.03$ and $1.30 \pm 0.07$. Compared to this
sample of galaxies, our clustering results are therefore somewhat lower.
This could be understood if the stellar masses of our simulated galaxies
are higher for a given space density than in the observations. 
Another possibility is that the relatively small box size of the simulation
(with the fundamental mode of the box going mildly non-linear at $z=1$) is
affecting the result. As we mentioned above, this should be less of a problem
for the prediction of the relative bias between dark matter and galaxies
than for the raw galaxy $\xi$, but should still be borne in mind.
We leave further investigation of the mass function by stellar 
mass of the simulated galaxies to future work.

The clustering of DEEP2 galaxies around  SDSS quasars has been measured by
Coil \etal (2007),
who find from the cross correlation that they are distributed
similarly, with no dependence on scale, quasar luminosity or redshift.
In the future, the {\it BHcosmo} simulation can be used to compare to
these observations in detail. For now we note that some
 qualitatively similar findings (limited dependence of quasar properties
on environment) has been found in the simulation  by Colberg \& Di Matteo
(2008).

\section{Summary and Discussion}

\subsection{Summary}

We have examined the galaxies forming in a cosmological 
hydrodynamic simulation 
 which includes self consistent modelling of black hole
growth and feedback. The mass and spatial resolution of the simulation
are similar to those of previous zoomed simulations of individual galaxies,
enabling us to explore to a similar accuracy the properties of
a population of $\sim1000$ galaxies with circular velocities
$\simgt 100 \kms$. 
Our conclusions can be summarised as follows.

\noindent(1) At redshift $z=1$ we find that galaxies can
be roughly divided into classes on the basis
of their morphology, kinematics and age. These can be
associated with the classic division into ``early'' and  ``late'' types,
so that on average most of the
usually assumed relationships between galaxy properties hold.
These relationships are that smaller galaxies tend to be
younger, have smaller Sersic indices, higher kinematic disk
to bulge ratios and star formation rates than older, larger galaxies.
 
\noindent(2) The angular momentum content of galaxies 
compared to observations is still a problem, as it is for the zoomed
simulations.  We find that disk galaxies in the simulation have 
on average $\sim 8$ times less specific angular momentum for a given
circular velocity than in the observational compilation of 
Navarro (1998).  
The kinematic decomposition of galaxies into disk and bulge fraction
also leads to a much higher proportion of bulge dominated galaxies
than in observations.
The inclusion of black hole physics does not appear
to alleviate the angular problem (also found in the zoomed simulations
of Okamoto \etal 2007). 

\noindent (3) The scale
length of disk galaxies for a given circular velocity
is approximately consistent with observations. The flattened
parts of galaxies are therefore partially supported by 
non-circular motions. This is related to the fact that the 
simulated disk galaxies also compare acceptably  to the 
$z=1$ stellar mass Tully-Fisher relation observations of Kassin \etal (2007),
who include a non-circular component (gas velocity dispersion) in their
preferred statistic. When computing the same statistic
for the simulations, the overall agreement is acceptable, but with a different 
ratio of rotational and non-rotational motions than the observations.
The extended stellar populations of galaxies are approximately the right size,
but are kinematically distributed differently. 

\noindent (4) The properties of galaxies 
in the simulation are related to their
large scale environment, in a fashion which changes with redshift.
At $z=0$, the higher fraction of early types in dense environments is
qualitatively similar to that seen in observational data (e.g.,
Dressler 1980). At higher redshifts, the observational and
simulation trends are both for relatively more late types
at high densities. However, the simulation predicts somewhat too many,
with an actual reversal of the morphology density relation.

\noindent(5) The overall fraction of disky (late type) galaxies
 in the simulation decreases as we move to lower redshifts, changing from
$>0.7$ at $z\sim10$ to $\sim 0.45$ at $z=0$.

\noindent(6) The clustering of galaxies
(with $v_{\rm circ}>100 \kms$) is slightly stronger
than that of the dark matter, with
the correlation function $\xi(r)$
exhibiting an average positive bias factor $b\sim1.2$
on scales $r>0.5 \hmpc$. At $z=1$,
we find that late type galaxies (selected
by either age, kinematics or Sersic index) are actually more positively biased
than early types, a reversal of the trend observed at low redshifts.

\subsection{Discussion}

As with previous zoomed simulations of disk galaxies, our results
are mixed. There are many aspects of the simulated
galaxies which agree qualitatively and quantitatively with observational
data, but there are also still important differences. The
large fraction of mass in bulge components, linked to the lack of
orbital stellar angular momentum are the most obvious. In order to
form more realistic disk galaxies in this type of simulation, it is likely
that subtleties in the feedback mechanisms need to be changed. Okamoto
\etal (2005) have investigated different models of feedback in zoomed 
simulations, finding that galaxy morphologies are extremely sensitive
to how the feedback is applied and how it is related to star formation.
For example, they find that when star formation is formed through shock
induced compression of gas (a ``burst'' mode), the feedback heats up
a large reservoir of gas which later cools to form an extended
rotationally supported disk. On the
other hand, Okamoto \etal (2005) find that concentrating
injection of feedback energy in high density regions  (a rough approximation
of   AGN feedback) does not have the same effect. We have seen in this
paper (as Okamoto \etal 2007 also did with zoomed simulations with black
holes) that modelling AGN feedback with black hole particles does indeed
not solve the angular momentum problem. Clearly, different models of feedback
should be explored in simulations which encompass a fair sample of galaxies.
In the present work, we have shown that the multiphase model for
star formation of Springel \& Hernquist (2003) can be used to simulate
galaxies in a uniform cosmological run with the same relative degree of
success as in the zoomed runs of  Robertson \etal (2004).

Numerical and algorithmic issues can also influence the morphology
of galaxies as well as implementations of feedback physics. For
example, Governato \etal (2007) have found that increasing numerical
resolution has led to some improvement in the rotational support of
disks and less transfer of angular momentum to the dark matter. We
have seen in the present paper that our low resolution D4 run
is not very much different that the {\it BHCosmo} run, so that it seems 
unlikely that straightforward decreases 
in particle mass and gravitational
softening length would result in
significantly better agreement with observations than
could be obtained without also changing the way star formation and 
feedback are modelled.  
In the present
work, we have made use of multiphase modelling of
subresolution star formation physics which Springel 
\& Hernquist (2003b) have shown exhibits good convergence properties.
However, better modelling of
star formation and less dependence on subresolution parametrization
would go hand in hand with increased resolution. For
example, Pelupessy \etal (2006) find that explicit modelling 
of molecular hydrogen formation in  SPH simulations of dwarf
galaxies requires gravitational softenings on the order of tens of pc
and gas particle masses of $\sim 10^{3} \msun$. Ceverino \& Klypin
(2007) have recently shown with AMR simulations of hot gas bubbles caused
by star formation that modelling these bubbles self consistently without
subresolution assumptions requires 50pc resolution. 
 
On the algorithmic side, there is the possiblity that issues with SPH
are affecting the gas physics that can be modelled well.
For example, Agertz \etal (2007) have recently
shown that spurious pressure forces on particles can arise in 
regions where there are steep density gradients
(but see Price, 2007  for natural ways to resolve this within the
SPH formalism).
 On the other hand,
Bryan (2007) has shown that the overly centrally concentrated 
density distributions and rotation
curves that are a generic feature of simulations of spiral
galaxies also occur in extremely high resolution simulations
of disks using Adaptive Mesh Refinement. Bryan (2007) concludes that 
more sophisticated modelling of the interstellar medium, including 
star formation and turbulent support is needed to resolve these issues
in galaxy formation.

Some of the results in this paper may have a slightly different
interpretation if the classification of early and late type galaxies
is broadened into more categories. For example,
dwarf ellipticals and their subtype dwarf spheroidals are known
to have many of the properties of spiral galaxies rather than larger
ellipticals (e.g., Grebel \etal 2003). 
It is possible that some of the effects we are seeing are
partly due to this (for example the higher clustering amplitude
of late types at high $z$ could partially be 
because there are many dwarf galaxies in the early type sample).
Our lower cutoff in mass corresponds to a circular velocity $\sim 100 \kms$,
probably too high for most true dwarfs, but still low enough that these
possiblities should be investigated further in future work. 
It is also possible that irregular galaxies and their properties could
be investigated separately. Future work at higher resolution
will eventually make this possible.

The presence of late-type galaxies at high redshifts
in high density environments which we
have seen in \S4.3 in the {\it BHCosmo} simulation has long been 
seen observationally. For example, clusters at $z>0.4$
contain substantial populations of blue spiral galaxies (e.g., Butcher
and Oemler 1978) which evolve much more rapidly than field
galaxies in the 4-5 Gyr timescale to $z=0$. Transformation of these
galaxies by ``galaxy harassment'' (Moore \etal 1996) into
dwarf ellipticals may be ongoing in our simulation, and 
partially account for the
rising fraction of early types at lower redshifts. We can see from the
morphology-density relation in Figure 13 that there is indeed much
more evolution of the galaxy population at the high density end.

Van den Bosch \etal  (2007) have investigated the morphological
tranformations that take place in high density environments in the SDSS,
looking at the difference between central and satellite galaxies. They find
that whenever a central galaxy falls into a bigger halo (i.e., becomes
a satellite), it undergoes a color change (becoming slightly redder)
and a  morphology change (becoming very slightly more concentrated).
They find that  the magnitude of this effect is independent
of the mass of the new host
halo in which the galaxy falls. This suggests that ram-pressure stripping
of cold gas is not the likely cause, but instead ``strangulation'',
or removal of the galaxy halo hot gas. This takes place over a 
relatively long timescale (not accounted for in
past  semi-analytic models,
as shown by Kang \& van den Bosch (2008). In our simulation, we do find
in Figure 12 that 
galaxies become more centrally concentrated ($n$ increases) as 
as a function of decreasing distance from the host halo center.
Tracking the role of gas accretion and loss from the galaxies (as carried out
by e.g., Kere\v{s} \etal 2005) could be used to investigate the
mechanism by which this occurs.

Much of the focus of galaxy formation simulations has been on trying to 
form realistic disks. We have seen in the present work that we are able
to achieve a similar degree of success (for example in the specific
angular momentum-circular velocity relationship) in our full volume
cosmological run to earlier zoomed simulations (e.g., Robertson \etal 2004,
Abadi \etal 2003ab). We have not looked in so much detail however at the
elliptical galaxies formed in the {\it BHCosmo}. Naab \etal (2007)
have recently performed zoomed simulations of early-type galaxies
from $\Lambda$CDM initial and find that realistic intermediate mass 
elliptical galaxies can be formed even without recent major mergers or
AGN and supernova feedback. We have mass resolution throughout our
box only $\sim 3$ times worse, than in the fiducial zoomed runs
of Naab \etal (although we have significantly coarser spatial
resolution), so that it would be worth investigating the properties of
the larger population of early types in more detail in future work.

In the future, it will be necessary to run a larger box down to
redshift $z=0$. In this way, it will be possible to investigate a 
wider range of environments, including rich clusters, where the evolutionary
history of simulated galaxies is expected to be different from
galaxy groups and the field. A much wider range of observational
data is also available at $z=0$. Population synthesis modelling of the
simulated galaxies would enable more direct comparisons with the data,
including the effect of observing in different wavelengths
on morphology. 

\section*{Acknowledgments}
We thank J\"{o}rg Colberg for providing the subhalo files and  J\"{o}rg
Colberg and Andrew Zentner for useful discussions.
 This work was partially supported by the
National Science Foundation, NSF AST-0205978 and NSF AST-0607819.
The simulations were carried out at Carnegie Mellon University and on the
Cray XT-3 at the Pittsburgh
Supercomputing Center.

\end{document}